\documentclass[letterpaper, 12pt]{article}
\usepackage[margin=1in]{geometry}
\usepackage{amsmath,amssymb,amsfonts,amsthm,bbm}
\usepackage{epic,eepic,epsfig,longtable}
\usepackage{multirow,verbatim}
\usepackage{array}
\usepackage{graphicx}
\usepackage{floatrow}
\usepackage{subcaption}
\usepackage{paralist}
\usepackage{latexsym}
\usepackage{comment}
\usepackage{epsfig}
\usepackage{setspace}
\usepackage{CJK}
\usepackage{color}
\usepackage{rotating}
\usepackage[authoryear,round]{natbib}
\makeatletter
\def\singlespace{\def\baselinestretch{1}\@normalsize}

\numberwithin{equation}{section}

\renewcommand{\hat}{\widehat}

\renewcommand{\hat}{\widehat}

\newcommand{\bfm}[1]{\ensuremath{\mathbf{#1}}}

    \def\FF{\mathbb{F}}

   \def\bU{\bfm U}

\newcommand{\bfsym}[1]{\ensuremath{\boldsymbol{#1}}}

\def\1{\bfsym{1}}	

\DeclareMathOperator{\E}{E}

\def\newpage{\vfill\eject}

\def\today{\ifcase\month\or
  January\or February\or March\or April\or May\or June\or
  July\or August\or September\or October\or November\or December\fi
  \space\number\day, \number\year}

\newdimen\biblioindent    \biblioindent=30pt

 at 8truept

\newcommand{\beq}{\begin{equation}}
  \newcommand{\eeq}{\end{equation}}
\newcommand{\beqn}{\begin{eqnarray}}
  \newcommand{\eeqn}{\end{eqnarray}}
\newcommand{\beqnn}{\begin{eqnarray*}}
  \newcommand{\eeqnn}{\end{eqnarray*}}

\allowdisplaybreaks
\usepackage{verbatim}
\def\tilde{\widetilde}

\def\FF{\mathcal{F}}
\def\[{\left [}  \def\]{\right ]} \def\({\left (}  \def\){\right )}
 \def\endpf{$\blacksquare$}
\def\hat{\widehat}

\newtheorem{assumption}{Assumption}
\newtheorem{theorem}{Theorem}
\newtheorem{lemma}{Lemma}

\newtheorem{proposition}{Proposition}
\theoremstyle{definition}

\newtheorem{remark}{Remark}

\title{Volatility Models for Stylized Facts of High-Frequency Financial Data}
\author{Donggyu Kim\footnote{Corresponding author. Address: College of Business, KAIST, Seoul 02455, South Korea. Phone: +82-02-958-3448. E-mail: dkim0329@gmail.com.}
  and Minseok Shin  \\
Korea Advanced Institute of Science and Technology (KAIST)
}

\begin{document}
\maketitle

\begin{spacing}{2}

\begin{abstract}
This paper introduces novel volatility diffusion models to account for the stylized facts of high-frequency financial data such as volatility clustering, intra-day U-shape, and leverage effect.
For example, the daily integrated volatility of the proposed volatility process has a realized GARCH structure with an asymmetric effect on log-returns.
To further explain the heavy-tailedness of the financial data, we assume that the log-returns have a finite $2b$-th moment for $b \in (1,2]$. 
Then, we propose a Huber regression estimator which has an optimal convergence rate of $n^{(1-b)/b}$.
We also discuss how to adjust bias coming from Huber loss and show its asymptotic properties. 
\end{abstract}



\section{Introduction} \label{SEC-1}
 
Volatility plays a pivotal role in financial applications such as portfolio allocation, risk management, and performance evaluation.
Low-frequency and high-frequency financial data are widely used to analyze volatility, and several stylized facts have been found. 
For example,  we often observe that large changes of returns tend to be followed by large changes, and small changes of returns tend to be followed by small changes, which is known as the volatility clustering \citep{mandelbrot1963variation}.
Several empirical studies have shown a negative correlation between financial return and future volatility, which is called the leverage effect \citep{ait2017estimation, black1976studies, christie1982stochastic, figlewski2000leverage, tauchen1996volume}. 
To explain these stylized facts, several discrete-time series models have been introduced.
For example, generalized autoregressive conditional heteroskedasticity (GARCH) has been introduced to capture volatility dynamics \citep{bollerslev1986generalized, engle1982autoregressive}.
To further explain the leverage effect,  several variations of the GARCH models have been proposed  with some asymmetric effects of stock returns, such as the  GJR-GARCH \citep{glosten1993relation},  NG-GARCH \citep{engle1993measuring}, and QR-GARCH \citep{nyberg2012risk} models. 
To account for more general state heterogeneity, Markov-switching GARCH models have been introduced \citep{bauwens2010theory, bauwens2014marginal, gray1996modeling, haas2004new,  hamilton1994autoregressive, klaassen2002improving}. 
However, when the volatility changes rapidly to a new level, it is often difficult to catch up with it immediately using only the daily log-returns as the innovations \citep{andersen2003modeling}.

On the other hand, several empirical studies show that incorporating high-frequency information helps to explain market dynamics \citep{chun2022state, corsi2009simple, hansen2012realized, kim2016statistical, kim2019factor, kim2016unified, shephard2010realising, song2021volatility}. 
For high-frequency information on volatility modeling, realized volatility estimators, such as multi-scale realized volatility (MSRV) \citep{zhang2006efficient, zhang2011estimating}, pre-averaging realized volatility (PRV) \citep{jacod2009microstructure}, quasi-maximum likelihood estimator (QMLE) \citep{ait2010high, xiu2010quasi}, kernel realized volatility (KRV) \citep{barndorff2008designing}, wavelet realized volatility \citep{fan2007multi}, robust pre-averaging realized volatility \citep{fan2018robust}, and adaptive robust pre-averaging realized volatility \citep{shin2021adaptive} have been proposed.
These estimators use the sub-sampling and local-averaging techniques to remove the effects of market micro-structure noises and estimate the integrated volatility consistently and efficiently. 
To further identify the jump locations given noisy high-frequency data, \citet{fan2007multi} and \citet{zhang2016jump} proposed wavelet methods to detect jumps and applied the MSRV method to jump-adjusted data. 
Also, the pre-averaging realized estimator can mitigate the jump variation by using thresholding scheme and estimate the jump variation consistently \citep{ait2016increased, jacod2009microstructure}.
With these realized volatility estimator as the new innovation, several time series models have been developed.
 Examples include  heterogeneous auto-regressive (HAR) models \citep{corsi2009simple},  high-frequency based volatility (HEAVY) models \citep{shephard2010realising}, realized GARCH models \citep{hansen2012realized}, unified GARCH-It\^o models \citep{kim2016unified}, realized GARCH-It\^o models \citep{song2021volatility}, and factor GARCH-It\^o models \citep{kim2019factor}.
Since realized volatility is based on  high-frequency data, the time series models above need to connect with continuous-time diffusion processes. 
When handling intra-day continuous-time series models, we often observe  an intra-day U-shape volatility pattern \citep{admati1988theory, andersen1997intraday, andersen2018time, hong2000trading} and price jumps \citep{ait2012testing,   andersen2007roughing, barndorff2006econometrics, corsi2010threshold}.   
However,  to our best knowledge, the existing continuous-time volatility models do not explain all of the well-known stylized facts observed in the inter-day and intra-day such as price jump, volatility clustering, leverage effect, and intra-day U-shape.  
Ignoring one of these stylized facts means failing to account for some market dynamics.
 This fact increases the demand of developing a diffusion process which can explain the well-known stylized facts of high-frequency financial data.

To analyze the asymptotic behaviors of volatility estimation procedures, we usually need some finite fourth-moment condition because the parameter of interest is the second moment. 
However, empirical studies have demonstrated that the bounded fourth-moment condition is often violated \citep{cont2001empirical, mao2018stochastic, massacci2017tail}.
To handle this heavy-tail problem,  \citet{sun2020adaptive} introduced the adaptive Huber regression for independent observations, which can obtain the optimal convergence rate with only the finite $b$-th moment for $b \in (1, 2]$. 
\citet{fan2019adaptive} extended the adaptive Huber regression to the dependent observations, such as Markov dependence. 
On the other hand, \citet{mikosch2006stable} investigated the stable limits of the usual Gaussian quasi-likelihood maximum estimator (QMLE) with the heavy-tailed observations. 
High-frequency financial data also suffer from the heavy-tailedness problem \citep{fan2018robust, shin2021adaptive}. 
Thus, the classical estimation method, which is used to estimate high-frequency time series models above, may be not  efficient neither effective. 
This fact urges us to investigate a robust estimation procedure in the volatility analysis based on high-frequency data.

In this paper, we develop  a novel volatility diffusion process to account for the stylized facts observed in the inter-day and intra-day, such as price jump, volatility clustering, leverage effect, intra-day U-shape, and heavy-tailedness.   
Specifically, to account for the intra-day U-shape,  instantaneous volatility has a quadratic form with respect to time.
For the volatility clustering, the magnitude of the instantaneous volatility is determined by the previous day's volatility, and for the leverage effect, the instantaneous volatility process has a negative correlation with the log-price process.  
To incorporate the high-frequency information as the innovation, the  increment  of the instantaneous volatility process has the current instantaneous volatility as the drift term. 
We further allow the price to have jumps. 
Then, we show that the daily integrated volatility of the instantaneous volatility has the well-known realized GARCH structure with an asymmetric effect of the log-returns. 
That is, the integrated volatility has an auto-regressive structure with additional daily log-returns, which helps to account for the low-frequency volatility dynamics such as the leverage effect and volatility clustering. 
We call it asymmetric realized GARCH-It\^o (ARGI).  
To model the heavy-tailedness, we assume that the log-return has only the finite $2b$-th moment for $b \in (1,2]$, so the total variation has only the finite $b$-th moment. 
For model parameter inference, we adopt total variation as a proxy for its conditional expectation and employ the adaptive Huber regression procedure \citep{sun2020adaptive}.
Then, we show its asymptotic normality with the optimal convergence rate $n^{(1-b)/b}$, and find that Huber loss causes a non-negligible bias.
To adjust for this bias, we propose a bias adjustment procedure and show that the bias adjusted estimator displays the Bahadur representation, which is the sum of the heavy-tailed martingale differences. 
We further discuss the stable limits of the Bahadur representation.

The rest of the paper is organized as follows. 
Section \ref{SEC-2}  introduces the ARGI model and investigate its property. 
Section \ref{SEC-3} proposes the adaptive Huber regression and establishes its asymptotic properties.
We also propose a bias adjusted estimator and investigate its asymptotic properties.
Section \ref{SEC-4}  conducts a simulation study to check finite sample performance.
In Section \ref{SEC-5}, we apply the proposed method to real high-frequency data and check its benefits. 
The conclusion is presented in Section \ref{SEC-6}. 
We collect the proofs in the Appendix.


\setcounter{equation}{0}
\section{Asymmetric realized GARCH-It\^o models} \label{SEC-2}

We assume that the stock log-price follows the jump diffusion process as follows:
\begin{equation}\label{jump-diffusion}  
	d X_t  = \mu  dt + \sigma_t dB_t+ J_t d L_t,
\end{equation}
where $\mu $ and $\sigma_t$ are  drift and instantaneous volatility processes, respectively, and $B_t$ is the independent standard Brownian motion. 
For the jump part, $J_t$ is an i.i.d. jump size with mean $\omega_L$, and $L_t$ is a independent Poisson process with  intensity $\lambda$.

In financial markets, we observe several stylized facts of the volatility process, such as the intra-day U-shape  pattern,  leverage effect, volatility clustering, and heavy-tail.
To reflect these stylized facts, we develop a novel instantaneous volatility process as follows.
We first consider the discrete increment process.
Let $\Delta f_t= f_{t+ \Delta} - f_{t}$ for a given process $f_t$. 
Then, for $t \geq 0$, we assume that the increment of the instantaneous volatility process has the following structure:
\begin{equation*} 
	\Delta \sigma_t ^2 = \{ 2 \gamma t (\omega_1+ \sigma_{\lfloor  t \rfloor}^2) - ( \omega_ 2 + \sigma_{\lfloor  t \rfloor}^2)   \} \Delta t + \beta   \sigma_t^2 \Delta t  - \alpha \Delta X_t + \varphi _t  \Delta W_t,  
\end{equation*}
where $\lfloor  t \rfloor$ is the integer part of $t$,  $W_t$ is a standard Brownian motion which can be correlated with $B_t$, $\varphi_t$ is a random fluctuation, and the model parameters $\omega_1, \omega_2, \beta, \alpha,$ and $\gamma$ are positive.  
For the deterministic part, the term $ 2 \gamma t (\omega_1+ \sigma_{\lfloor  t \rfloor}^2) - ( \omega_ 2 + \sigma_{\lfloor  t \rfloor}^2) $ can explain the intra-day U-shape  pattern by choosing appropriate parameters, and the previous period's volatility $\sigma_{\lfloor  t \rfloor}^2$ determines the magnitude of the current period, which is able to explain the volatility clustering.
The increment $- \alpha \Delta X_t$ from the price process has an asymmetric shock structure, which makes it possible to account for the leverage effect. 
In contrast, the $\beta   \sigma_t^2 \Delta t$ term tends to increase as the magnitude of $\Delta X_t$ increases, which provides the high-frequency information-based innovation term, integrated volatility.  
Finally, $\varphi _t $ represents random fluctuation.  
As $\Delta$ goes to zero, we have the following diffusion process:
\begin{equation}\label{diffu} 
	d \sigma_t ^2 = \{ 2 \gamma t (\omega_1+ \sigma_{\lfloor  t \rfloor}^2) - ( \omega_ 2 + \sigma_{\lfloor  t \rfloor}^2)   \} d t + \beta  \sigma_t ^2 dt  - \alpha d X_t + \varphi _t  d W_t. 
\end{equation}
Thus, for any $t\geq 0$, the instantaneous volatility process has the following solution:
\begin{eqnarray}\label{ARGI}  
\sigma_t ^2 & =&  \sigma_{\lfloor  t \rfloor}^2 + \gamma (t- \lfloor  t \rfloor) ^2 ( \omega_1 + \sigma_{\lfloor  t \rfloor}^2 ) - (t-\lfloor  t \rfloor) ( \omega_ 2 + \sigma_{\lfloor  t \rfloor}^2 ) + \beta  \int_{\lfloor  t \rfloor}^t \sigma_s ^2 ds  \cr
	&& - \alpha (X_t - X_{\lfloor  t \rfloor}) +  \int_{\lfloor  t \rfloor}^t \varphi_s dW_s \, \text{ a.s.}  
\end{eqnarray}
\begin{remark}\label{remark1}
The diffusion process \eqref{diffu} can generate negative paths for the instantaneous volatility process $\sigma_{t}^{2}$.
 However, by choosing appropriate model structure, we can mitigate the effect of the price jump and random fluctuation on the sign of $\sigma_{t}^{2}$. 
For example, let $n-1 \leq t < n $ for some positive constant $n$.
When the jump does not exist, by It\^o's lemma, we have
\begin{eqnarray*}  
&& \beta  \int_{n-1}^t \sigma_s ^2 ds - \alpha (X_t - X_{n-1}) \cr
&& = \beta \(X_t - X_{n-1} - \dfrac{\alpha}{2\beta}\)^2 -2\beta \int_{n-1}^t \(X_s - X_{n-1}\)dX_s -\dfrac{\alpha^2}{4\beta} \, \text{ a.s.}  
\end{eqnarray*}
Thus, by choosing appropriate model parameters and random fluctuation, we can obtain the positive instantaneous volatility process.
For example,  we can choose $\int_{n-1}^t \varphi_s dW_s = 2\beta \int_{n-1}^t \(X_s - X_{n-1}\)dX_s + c \int_{n-1}^t  (\tilde{W}_t -\tilde{W}_{n-1})  d \tilde{W}_t $, where $\tilde{W}_t$ is an independent standard Brownian motion. 
Then, by It\^o's lemma,  we have $\int_{n-1}^t \varphi_s dW_s -2\beta \int_{n-1}^t \(X_s - X_{n-1}\)dX_s  \geq  - c/2$ a.s.; thus,  by choosing appropriate parameters, we can obtain strictly positive instantaneous volatility process. 
On the other hand, when we allow the jumps, $\sigma_{t}^{2}$ does not have negative values with high probability under the condition that the positive jump size is bounded. 
We note that the size of the positive jump is often smaller than that of the negative jump \citep{jiang2013stock}.
Furthermore, when the number of positive jumps is bounded, $\sigma_{t}^{2}$ can have a positive path with probability one.
\end{remark}

To evaluate the proposed process, we need the initial value condition for each low-frequency period. 
In this model, the random fluctuation part $ \int_{n-1}^n \varphi_t dW_t$ is imposed to explain the random fluctuation of the volatility process, so it is purely random and does not contain the market dynamics information.  
In this point of view,   we assume that the random fluctuation part $ \int_{n-1}^n \varphi_t dW_t$  does not affect on the initial value of the instantaneous volatility for each day, and thus, the instantaneous volatilities at the integer time points satisfy: 
\begin{equation*}
	\sigma_n ^2   =  \omega+ \gamma \sigma_{n-1}^2 +  \beta \int_{n-1}^n \sigma_s ^2 ds - \alpha (X_n - X_{n-1}),
\end{equation*}
where $\omega=\gamma \omega_1- \omega_2$. 
Then, the proposed instantaneous volatility process is c\`adl\`ag.
For example, we can choose the randomness term as $\(1-t+\lfloor  t \rfloor\)   \int_{ \lfloor  t \rfloor }^t \nu (W_s - W_{  \lfloor  t \rfloor })  dW_s $.
We call this diffusion process the asymmetric realized GARCH-It\^o (ARGI) model. 


In the following proposition, we investigate the low-frequency time series structure  of the total variations, which will be used for statistical inferences. 
 
 \begin{proposition}\label{prop-integratedVol}  
  We have 
\begin{equation*}
	V_n = h_n (\theta) + D_n  \text{ a.s.},
\end{equation*}
  where  
         \begin{eqnarray*}
  			&&V_n  = \int_{n-1}^{n} \sigma^2_{s}   ds+ \int_{n-1}^{n} J_s^2 dL_s, \cr
 			&&h_n   (\theta) =\omega^g + \gamma h_{n-1}   (\theta)   + \beta^g \int_{n-2}^{n-1} \sigma^2_{s}   ds   - \alpha^g (X_{n-1} - X_{n-2}),
			\end{eqnarray*}
			\begin{eqnarray}
			\label{GARCH-parameter}
			&&  \theta = \left(\omega^g, \gamma,  \alpha^g, \beta^g \right), \quad  \alpha^g=\left( \rho_1 -\rho_2 +2\gamma \varrho_3 \right) \alpha  , \quad  \beta^g=  \left( \rho_1 -\rho_2 +2\gamma \varrho_3 \right)  \beta,     \cr 
			&& \omega^g = \gamma (\rho_1 -\varrho_2 + 2 \varrho_3) \omega_1 - ( \varrho_1 -  \gamma \varrho_2 +2\gamma \varrho_3 ) \omega_2 -  (1-\gamma)     \[   \varrho_2  \alpha ( \mu + \lambda   \omega_L )   -  \lambda E( J_t^2 )  \] , \cr
			&&  \rho_1 = \beta^{-1} (e^{\beta}-1), \quad \rho_2 = \beta^{-2} (e^{\beta}-1-\beta),  \quad \rho_3=\beta^{-3} (e^{\beta}-1-\beta-\frac{\beta^2}{2}),
			\end{eqnarray}
  			and
			\begin{eqnarray*}
			&&D_n = D_n^c +D_n^J, \cr
			&& D_n^c=   \beta^{-1}  \int_{n-1}^{n}  \(e ^{\beta (n-t)} -1 \)  \varphi_t  dW_t  - \alpha   \beta^{-1}  \int_{n-1}^{n}  \(e ^{\beta (n-t)} -1 \)  \sigma_t d B_t ,    \cr
			&& D_n^J =    \int_{n-1}^{n} \[ J_t ^2 - E \( J_t^2 \)      - \alpha  \beta^{-1} \(e ^{\beta (n-t)} -1 \) (J_t - \omega_L) \]  dL_t   \cr
				&& \qquad \qquad + \int_{n-1}^{n} \[ E \( J_t^2 \)  - \alpha  \beta^{-1} \omega_L \(e ^{\beta (n-t)} -1 \) \] (d L_t -\lambda dt )  
			\end{eqnarray*}
		are all martingale differences.
 \end{proposition}
 
Proposition \ref{prop-integratedVol} shows that the total variation, $V_n$, for each day is decomposed into the $\FF_{n-1}$-measurable $h_n(\theta)$ and martingale difference $D_n$.  
By It\^o's isometry, the conditional variance of the daily log-return is 
\begin{equation*}
	E \[ (X_n - X_{n-1} - \mu )^2 \middle | \FF_{n-1}  \] = h_n(\theta)\, \text{ a.s.}
\end{equation*}
Thus, $h_n(\theta)$ is the conditional variance of the daily log-return, which has the well-known GARCH structure with  two innovations: integrated volatility and daily log-return.  
This structure is able to catch the volatility clustering and leverage effect.

Since the intra-day dynamic structure, such as the intra-day U-shape, is averaged out,  the specific diffusion process defined in the ARGI model does not affect on the statistical inference for the low-frequency dynamics. 
That is, we only need the low-frequency time series structure derived in Proposition \ref{prop-integratedVol}:
\begin{equation} \label{V-structure}
	V_n = h_n (\theta) + D_n\,  \text{ a.s.}
\end{equation}
We note that any intra-day structure does not affect  the low-frequency dynamics as long as it is averaged out.  
Thus, for statistical inference for the low-frequency dynamics, we only assume  \eqref{V-structure}. 
On the other hand, to account for the severe heavy-tailedness such as the unbounded fourth moment, we impose heavy-tailedness on the jump size that has only the finite $2b$-th moment for $b \in (1, 2]$. 
Consequently, the log-return $X_n- X_{n-1}$ and the martingale difference $D_n$  have only the finite $2b$-th and $b$-th moments, respectively. 
Under this severe heavy-tailedness, the classical statistical estimation procedure is not effective nor efficient \citep{sun2020adaptive}. 
In the following section, we propose robust estimation procedures and establish their statistical properties.

 \section{Estimation procedure}  \label{SEC-3}
  
 \subsection{A model setup} \label{SEC-3.1}
 We assume that the underlying diffusion process follows the ARGI process defined in Section \ref{SEC-2}. 
High-frequency observations during the $d$-th day are observed at $t_{d,i}, i=1,\ldots, m_d$, where $d-1=t_{d,0} < t_{d,1} < \cdots < t_{d, m_d} = d$.  
Let $m$ be the average number of high-frequency observations, that is, $m= \frac{1}{n} \sum_{d=1}^n m_d$.  
The true prices $X_d$'s at integer times are observed.
By contrast,  due to market inefficiencies such as asymmetric information, bid-ask spread, and so on, high-frequency data are contaminated by micro-structure noises.
To account for this, we assume that the observed log-prices have the following additive noise structure:
 \begin{equation*}
 	Y_{t_{d,i}}= X_{t_{d,i}} + \epsilon_{t_{d,i}}, \quad \text{for } d=1,\ldots, n, i=1,\ldots, m_d-1, 
 \end{equation*}
where $X_t$ is the true log-price, and  $\epsilon_{t_{d,i}}$'s are micro-structure noises with mean zero and variance $\eta_d$.

With the noises and jumps, we first employ the non-parametric volatility estimator, which can be any noise- and jump- robust non-parametric volatility estimator \citep{ait2016increased, fan2007multi, jacod2009microstructure, shin2021adaptive, zhang2016jump}
In this paper,  for the $d$th day, we let $\hat{V}_d$, $JV_d$, and $RV_d$ be the estimators of the daily total variation,  jump variation, and  integrated volatility, respectively. 
Due to the heavy-tailedness of the jumps, the daily total variation estimator $\hat{V}_d$ is also heavy-tailed, which is decomposed into the true daily total variation $V_d$ and estimation error.
The estimation error can be decomposed into the martingale difference and some negligible non-martingale term, which is mainly coming from the drift term. 
Thus, the non-martingale term does not affect on the theoretical analysis, so for simplicity, we assume that the estimation error has only martingale difference which has  heavy-tail.  
Thus, we have
 \begin{equation}\label{model-vhat}
 	\hat{V}_d = h_d (\theta_0) + \tilde{D}_d,
 \end{equation}
 where $\theta_0$ is the true model parameter, and $\tilde{D}_d$ is the sum of  $D_d$ and estimation error, which has zero mean and the finite $b$-th moment.

 \subsection{Model parameters estimation}\label{SEC-3.2}

We first define some notations.
For any given  $d_1 \times d_2$ matrix $\bU = \left(U_{ij}\right)$,
denote its matrix spectral norm by $\|\bU\|_2$, its Frobenius norm $\|\bU\|_F = \sqrt{ \mathrm{tr}(\bU^{\top} \bU) }$,
 and
$    \| \bU \| _{\max} = \max_{i,j} | U_{ij}| $.
Let $C$'s be positive generic constants whose values are free of $\theta$, $n$, and $m_i$, and may change from appearance to appearance.

Proposition \ref{prop-integratedVol}  shows the linear relationship between the daily total variation $V_d$ and conditional variation $h_d(\theta_0)$.
Thus, it is natural to use the ordinary least squares (OLS) estimation procedure.
However,  due to the heavy-tailedness, the squared loss function does not work  \citep{sun2020adaptive}.
To handle the heavy-tailedness, we use the following Huber regression,  
  \begin{equation*}
 \hat{L} _{n,m} (\theta) 	= -\frac{1}{2n}  \sum_{i=1}^n  \ell_{\tau_n}  ( \hat{V}_i  - \hat{h} _ i (\theta) ),
 \end{equation*}
 where  $\tau_n= c_{\tau} n^{1/b}$ is a truncation parameter for some positive constant $c_{\tau}$, $\ell_{a} (\cdot)$ is the Huber loss 
\begin{eqnarray}\label{h_hat} 
&&\ell_ a  (x)= \begin{cases}
  2 a |x| -a^2 & \text{ if }  |x| \geq a \\ 
 x^2 & \text{ if }  |x| < a, \quad \text{and}
\end{cases} \cr
&&  \hat{h}_{ i} (\theta) =\omega^g + \gamma \hat{h}_{i-1}    (\theta)   + \beta^g RV_{i-1}    - \alpha^g (X_{i-1} - X_{i-2}). 
\end{eqnarray}
 Then, we estimate the model parameter $\theta_0$ by maximizing the quasi-likelihood function $ \hat{L} _{n,m} (\theta)$ as follows:
\begin{equation*}
	\hat{\theta}= \arg \max _{\theta \in \Theta} \hat{L}_{n,m} (\theta),
\end{equation*}
where $\Theta$ is the parameter space of $\theta$.

To investigate the Huber estimator $\hat{\theta}$, we need the following technical conditions. 
\begin{assumption} \label{assumption1} ~
	\begin{enumerate}
		\item[(a)] Let
		\begin{equation*}
		\Theta = \lbrace  (\omega^g, \gamma, \alpha^g, \beta^g): \omega_l^g < \omega^g < \omega_u^g, \gamma_l < \gamma < \gamma_u, \alpha_l^g < \alpha^g < \alpha_u^g, \beta_l^g <\beta^g < \beta_u^g,   \gamma + \beta^g <1  \rbrace, 
		\end{equation*}
		where $\omega_l^g, \omega_u^g, \gamma_l, \gamma_u, \alpha_l^g, \alpha_u^g, \beta_l^g, \beta_u^g$ are known positive constants.
		
		\item[(b)]  $\sup_{t} E \[ \sigma_t^{2b} \] \leq C$,  $ \sup_{i \in \mathbb{N} } E\[ |X_i - X_{i-1}  |^{2b} \] \leq C$ and   $\sup_{i \in \mathbb{N} } E \[ |\tilde{D}_i |^b\middle | \FF_{i-1} \] \leq C$ a.s. for some $b \in (1, 2]$.
		
		\item[(c)] There exist  fixed constants $C_1$ and $C_2$ such that $C_1 m \leq m_i \leq C_2 m$  and $\sup_{1 \leq j \leq m_i} | t_{i,j} - t_{i, j-1} | = O(m^{-1})$, and $n^{4(b-1)/b} m^{-1} \rightarrow 0$ as $m,n \rightarrow \infty$ .
  	
		\item[(d)] $\sup_{i\in \mathbb{N}}   | RV_{i}-\int_{i-1}^{i}\sigma_{s}^{2}ds |= O_p ( m^{-1/4} )$  .
 
		\item[(e)]   $ \tau_n ^{b-2} \[  \ell_{\tau_n}  ( \hat{V}_{i}- h_i(\theta)  )  - E \left \{ \ell_{\tau_n}  ( \hat{V}_{i}- h_i(\theta) )  \middle | \FF_{i-1}  \right \} \] $ is uniformly integrable.
		
		\item[(f)] $\left(\tilde{D}_i, D_i,  \int_{i-1}^{i} J_s^2 dL_s,  X_i - X_{i-1}  \right)$ is a stationary ergodic process.
	\end{enumerate}
\end{assumption}

\begin{remark} 
The moment condition Assumption \ref{assumption1}(b) represents for the heavy-tailedness for the log-prices and total variation. 
Assumption \ref{assumption1}(d) is required to handle the long period model set-up. 
Under the locally bounded condition for the instantaneous volatility process, this condition can be shown with the additional $\log n$ term \citep{fan2018robust, Kim2016SPCA}. 
We note that the heavy-tailedness of the price jumps does not affect on the convergence rate of the realized volatility estimator \citep{shin2021adaptive}. 
The uniformly integrable condition Assumption \ref{assumption1}(e) is a minimum requirement for investigating the consistency. 
We can show that the expectation of $\ell_{\tau_n}  (\hat{V}_{i}- h_i(\theta))$ is bounded by $\tau_n ^{2-b}$. 
Thus, the expectation of $\tau_n ^{b-2} \left|  \ell_{\tau_n}  ( \hat{V}_{i}- h_i(\theta) )  - E \left \{ \ell_{\tau_n}  ( \hat{V}_{i}- h_i(\theta) )  \middle | \FF_{i-1}  \right \} \right|$ is also bounded.
In this point of view, the uniformly integrable condition is not strong at all.
Finally, the stationary ergodic condition Assumption \ref{assumption1}(f) is required to obtain the asymptotic normality, and the condition for the daily return is required due to the asymmetric effect term. 
\end{remark}

\begin{remark}
As discussed in Remark \ref{remark1}, the instantaneous volatility process can have a positive path with probability one, which implies that the total variation $V_n$ and its conditional expectation $h_n(\theta_0)$ also have positive paths.
Thus, for an appropriate model parameter space,  $h_n(\theta)$ can always be positive.
Then, by Assumption \ref{assumption1}(d), we can guarantee the positiveness of $\hat{h}_i(\theta)$ with high probability.
In practice, we restrict the parameter space such that $\hat{h}_i(\theta) \geq 0$.
\end{remark}

The following theorem establishes the asymptotic behaviors of the Huber estimator $\hat{\theta}$.

\begin{theorem}\label{Thm-theta}
Under the model set-up in Section \ref{SEC-2} and Assumption \ref{assumption1}(a)--(e) (except for $n^{4(b-1)/b} m^{-1} \rightarrow 0$),  we have 
 \begin{equation}\label{Thm-theta-r1}
  \| \hat{\theta} - \theta_0 \| _{\max} = O_p (m^{-1/4} + n^{(1-b)/b}  ) .
 \end{equation}
Furthermore,  if Assumption \ref{assumption1} is met and $b\in (3/2, 2]$, we have
\begin{equation}\label{Thm-theta-r2}
n^{(b-1)/b}  ( \hat{\theta} - \theta_0) \overset{d}{\to} N(    V_2 ^{-1} S, \,  V_1 V_2 ^{-1} ),
 \end{equation}
where 
 \begin{eqnarray*}
&& V_1 =   \lim_{n \to \infty}   n ^{(b-2)/b}  E \[     \tilde{D}_i ^2     \1_{\{ | \tilde{D}_i |  \leq  \tau_n \} }     +    \tau_n   ^2   \1_{\{| \tilde{D}_i |  \geq  \tau_n \}}      \]  ,  \quad  V_2= E \[  \frac{\partial h_i   (\theta_0) }{\partial \theta}    \frac{\partial h_i   (\theta_0) }{\partial \theta ^{\top} }  \]  , \cr
 && S = \lim_{n \to \infty}  n^{(b-1)/b}   E \[   \left \{ -   \tilde{D}_i    \1_{\{ | \tilde{D}_i   | \geq \tau_n \} }  +   \tau_n  \1_{\{  \tilde{D}_i   \geq \tau_n \} }   -  \tau_n  \1_{\{\tilde{D}_i   \leq  -\tau_n \} }    \right \}  \frac{\partial h_i  (\theta_0) }{\partial \theta}  \]  .
 \end{eqnarray*}
\end{theorem} 

\begin{remark}
Theorem \ref{Thm-theta} indicates that the Huber estimator $\hat{\theta}$ has the convergence rate $m^{-1/4}$ and $n^{(1-b)/b}$.
The $m^{-1/4}$ term is the cost of estimating the integrated volatility. 
The $n^{(1-b)/b}$ term is the optimal convergence rate with the heavy-tailed observations \citep{sun2020adaptive}. 
Theorem \ref{Thm-theta} shows that the Huber estimator $\hat{\theta}$ has the classical asymptotic normality.
\end{remark}

The Huber regression estimation is biased due to the truncation method. We note that the optimal rate can be obtained when the squared bias and variance terms have the same rate.
Hence, the bias term is not negligible unless the martingale difference term $\tilde{D}_d$ is symmetric. 
However, in finance practice, we often observe the skewed observations, so $\tilde{D}_d$ may not be symmetric. 
To obtain the unbiased estimator, we need to adjust the bias term.
We use the  bias adjustment scheme as follows:
\begin{equation*}
\hat{\theta}_ {adj} = \hat{\theta} +    \[     \sum_{i=1}^n   \frac{\partial \hat{h}_i ( \hat{ \theta} ) }{\partial \theta} \frac{\partial \hat{h}_i ( \hat{\theta} ) }{\partial \theta ^{\top} } \] ^{-1}   \sum_{i=1}^n   \frac{\partial \hat{h}_i   (\hat{\theta}) }{\partial \theta} (\hat{V}_i - \hat{h}_i (\hat{\theta}) ).  
\end{equation*}
We call this the Adjusted-Huber estimator.
The following theorem investigates the asymptotic behavior of the Adjusted-Huber estimator $\hat{\theta}_ {adj}$.

\begin{theorem}\label{Thm-adj}
Under the model set-up in Section \ref{SEC-2} and Assumption \ref{assumption1}, we have
\begin{equation*}
 \hat{\theta}_ {adj} - \theta_0=  \frac{ V_2 ^{-1} }{n} \sum_{i=1}^n   \frac{\partial  h_i   (\theta_0) }{\partial \theta}  \tilde{D}_i +   o_p( \hat{\theta} -\theta_0 ) + O_p(m^{-1/4}).
\end{equation*}
\end{theorem}

Theorem \ref{Thm-adj} shows that $\hat{\theta}_ {adj} - \theta_0$ is decomposed into the Bahadur representation term and the negligible term. 
The Bahadur representation term $\frac{ V_2 ^{-1} }{n} \sum_{i=1}^n   \frac{\partial  h_i   (\theta_0) }{\partial \theta}  \tilde{D}_i $ is a mean zero process.
Thus, unlike the Huber estimator $\hat{\theta}$, the bias adjusted estimator is unbiased. 
However, $\tilde{D}_i$ is heavy-tailed, so it does not have the classical asymptotic normality result. 
To analyze its asymptotic distribution, we need some stable distribution assumption \citep{mikosch2006stable}.
For example, we assume that   $\frac{\partial  h_i   (\theta_0) }{\partial \theta}  \tilde{D}_i $ is $\beta$-mixing and $b$-stable.
By the Markov's inequality, we can show $P ( |\tilde{D}| \geq n^{1/b} )  \leq C n^{-1}$,  so we can find $a_n= c_a n^{1/b}$ for some constant $c_a$ such that $nP( |\tilde{D}| \geq a_n)\to 1$ as $n$ goes to infinity. 
Then, by Theorem 3.2 \citep{mikosch2006stable}, we have
\begin{equation*}
	a_{n}^{-1} \sum_{i=1}^n   \frac{\partial  h_i   (\theta_0) }{\partial \theta}  \tilde{D}_i \overset{d}{\to} N_{b},
\end{equation*}
where $N_b$ is a $b$-stable random vector.
Thus, we have
\begin{equation}\label{stable}
	n^{(b-1)/b} (\hat{\theta}_ {adj} - \theta_0)\overset{d}{\to}  c_{a}  V_2 ^{-1}  N_{b}.
\end{equation}
 When $b = 2$, the stable random vector is a normal random vector.

\subsection{A selection of the tuning parameter}\label{SEC-3.3}

To evaluate the Huber loss regression, we need to choose the truncation parameter $\tau_n$. 
We first discuss the tuning parameter $b$. 
To estimate $b$, we use the Hill's estimator \citep{hill1975simple} $\hat{\upsilon}$  as follows:
\begin{equation}\label{Hill}
	\hat{b}= \max\left(c_b, \min\left(2, \hat{\upsilon} \right)\right) \quad \text{and} \quad 	\hat{\upsilon}= \[\frac{1}{k_n} \sum_{i=0}^{k_n-1} ( \log \hat{V}_{(n-i)} - \log \hat{V}_{(n-k_n +1)})  \]^{-1},
\end{equation}
where $\hat{V}_{(k)}$ is the $k$-th order statistics of $\hat{V}_1, \ldots, \hat{V}_n$, and $k_n$ diverges with $k_n / n \to 0$ as $n$ goes to infinity.  
We choose $k_n$ as $\lfloor 4n^{1/2} \rfloor$ and $c_b=1.1$.   
Also, we choose the constant part $c_{\tau}$ as follows:
\begin{equation}\label{Threshold}
	c_{\tau}=  \frac{c}{n} \sum_{i=1}^n \left|\hat{V}_i - \bar{V}\right|^{\hat{b}}, 
\end{equation}
where $\bar{V}= \frac{1}{n} \sum_{i=1}^n \hat{V}_i$. In the empirical study, we set $c$ as 0.2.

\begin{remark}
The choice of  $c_b$ is related with the highest moment of the log-returns, such as the $2c_b$-th moment. 
 In this paper, since we handle the second moment of the log-returns, we need at least second moment. 
Thus, we use $c_b=1.1$, which, in empirical study, provides the robust empirical result. 
For the choice of $c$, in the numerical perspective, large $c$ and small $c$ lead to the large variance and large bias, respectively.
We find that choosing $c=0.2$ gives appropriate bias and variance, and the performance do not significantly depend on the choice of $c$ around $0.2$.
Finally, for $k_n$, under some regularity condition, we can show its consistency for the dependent observations \citep{hill2010tail, hsing1991tail}.
However, we find that small $k_n$ leads to the unstable results, so we chose $k_n = 4\sqrt{n}$. 
Unfortunately, the choice of tuning parameters is based on the numerical study.
It would be interesting and important to develop a optimal tuning parameter choice procedure that can enjoy both practically and theoretically good properties.
We leave this issue for future studies.
\end{remark}

\section{A simulation study}  \label{SEC-4}
In this section, we conducted simulations to check the finite sample performance of the proposed estimation methods.
The data were generated for $n$ days with frequency $1/m^{all}$ on each day and let $t_{d,j}=d-1+j/m^{all}, d=1, \ldots, n, j=0, \ldots, m^{all}$.
The true log-price follows the ARGI model:
\begin{equation*}
	d X_t  = \mu  dt + \sigma_t dB_t+ J_t d L_t,
\end{equation*}
where the instantaneous volatility process $\sigma_{t}^2$ satisfies
\begin{eqnarray*}  
\sigma_t ^2 & =&  \sigma_{\lfloor  t \rfloor}^2 + \gamma (t- \lfloor  t \rfloor) ^2 ( \omega_1 + \sigma_{\lfloor  t \rfloor}^2 ) - (t-\lfloor  t \rfloor) ( \omega_ 2 + \sigma_{\lfloor  t \rfloor}^2 ) + \beta  \int_{\lfloor  t \rfloor}^t \sigma_s ^2 ds  \cr
	&& - \alpha (X_t - X_{\lfloor  t \rfloor}) +  \(1-t+\lfloor  t \rfloor\)   \int_{ \lfloor  t \rfloor }^t \nu (W_s - W_{  \lfloor  t \rfloor })  dW_s  \, \text{ a.s.},
\end{eqnarray*}
the drift $\mu=0.02$, $\nu=0.01$, and $\left(\omega_1, \omega_2, \gamma, \alpha, \beta \right)=\left(3.9527, 0.1000, 0.2474, 0.3972, 0.2939 \right)$.
We take $X_0=10$ and $\sigma_{0}^{2}=\E\left(\sigma_{1}^{2}\right)=1.7462$.
For the jump part, we set intensity $\lambda=20$ and the mean zero i.i.d jump size $J_t$ follows  
\begin{eqnarray*}
	J_t=
 \begin{cases}
  -c_J \times \left(t_{df,t}\right)^2  & \text{ with probability 0.5 }    \\
 c_J \times Z_{t}^2 & \text{ with probability 0.5, } \\
\end{cases}
\end{eqnarray*}
where $t_{df,t}$ is an independent t-distribution with degrees of freedom $df$ and standard deviation $1$, and $Z_t$ is an independent standard normal distribution. 
We chose $c_J=0.04$ and $df=6$.
We note that the distribution of $J_t$ represents the asymmetry and heavy-tailedness of the financial data. 
Then, the target ARGI model parameter is  $\theta = \left(\omega^g, \gamma,  \alpha^g, \beta^g \right)= \left(0.9412, 0.2474, 0.2774, 0.2053 \right)$. 
We note that, in the empirical study, the A-Hub estimates for the last sample period are $\omega^{g}=5.883 \times 10^{-6}$, $\gamma=2.474 \times 10^{-1}$, $\alpha^{g}=6.937 \times 10^{-4}$, and $\beta^{g}=2.225 \times 10^{-1}$.
Thus, we used the same $\gamma$ and $\beta^{g}$, and $\omega^{g}$ was scaled by 160000 times for scale-up.
For $\alpha^{g}$,  to reflect the scale relationship, it was scaled by $\sqrt{160000}=400$ times.
The noise-contaminated observations were generated as follows:
\begin{equation*}
	Y_{t_{d,j}}  = X_{t_{d,j}} + \epsilon_{t_{d,j}} \text{ for } d=1, \ldots, n, j=1, \ldots, m-1,
\end{equation*}
where $m$ is the number of high-frequency observations for one day, and
$\epsilon_{t_{d,j}}$’s are generated from i.i.d normal distributions with mean $0$ and standard deviation $0.01$. To generate the true process, we chose $m^{all} = 23400$, and we varied $m$ from $390$ to $23400$ and $n$ from $125$ to $500$.
To estimate the total variation and integrated volatility, we utilized the jump adjusted
pre-averaging realized volatility (PRV) estimator \citep{ait2016increased, jacod2009microstructure} as follows:
\begin{equation}\label{PRV-total}
	\hat{V}_{d}=\frac{1}{\psi K} \sum^{m-K+1}_{k=1}\left\{\bar{Y}^2\left(t_{d,k}\right)-\frac{1}{2}\hat{Y}^2\left(t_{d,k}\right) \right\},
\end{equation}
\begin{equation}\label{PRV-integratedvol}
	RV_{d}=\frac{1}{\psi K} \sum^{m-K+1}_{k=1}\left\{\bar{Y}^2\left(t_{d,k}\right)-\frac{1}{2}\hat{Y}^2\left(t_{d,k}\right) \right\}\mathbf{1}\left\{\left| \bar{Y}\(t_{d,k}\) \right|\leq \varpi_{m}\right\},
\end{equation}
where
\begin{equation*}
	\bar{Y}\left(t_{d,k}\right)=\sum^{K-1}_{l=1}g\left(\frac{l}{K}\right)\left(Y_{t_{d,k+l}}-Y_{t_{d,k+l-1}}\right), 
	\quad \psi=\int_0^1 g\left(t\right)^2 dt,
\end{equation*}
\begin{equation*}
	\hat{Y}^2\left(t_{d,k}\right)=\sum^{K}_{l=1}\left\{g\left(\frac{l}{K}\right) - g\left(\frac{l-1}{K}\right) \right\}^2
	\left(Y_{t_{d,k+l-1}}-Y_{t_{d,k+l-2}}\right)^2,
\end{equation*}
$\mathbf{1}\left\{\cdot \right\}$ is the indicator function, and $\varpi_{m} = c_{\varpi}m^{-0.235}$ is a truncation parameter for some constant $c_{\varpi}$.
We chose the bandwidth parameter $K = \lfloor m^{1/2} \rfloor$,
weight function $g\left(x\right) = x \wedge \left(1 - x\right)$, and
$c_{\varpi}$ as 7 times the sample standard deviation for the pre-averaged variables $m^{1/8}\bar{Y}\left(t_{d,k}\right)$.
To estimate $b$, we employed the tuning parameter selection method \eqref{Hill}--\eqref{Threshold} in Section \ref{SEC-3.3}.
For comparison, we employed the following quasi-maximum likelihood estimator (QMLE):
\begin{equation*}
	\hat{\theta}= \arg \min _{\theta \in \Theta} \sum_{i=1}^{n}\dfrac{\hat{V}_i}{\hat{h}_i\(\theta\)}+ \log \hat{h}_i(\theta),
\end{equation*} 
where $\hat{h}_i\(\theta\)$ is defined in \eqref{h_hat}. 
We repeated the whole simulation procedure 1000 times.

\begin{sidewaystable}[htbp]
\caption{The squared biases, variances, and  MSEs  for the OLS, Huber, Adjusted-Huber, and QMLE methods of estimating ARGI model parameters for $n=125, 250, 500$ and $m=390, 780, 2340, 23400$. 
Note that Hub and Adj represent the Huber and Adjusted-Huber methods, respectively.}\label{Table-1}
\centering
\scalebox{0.56}{
\begin{tabular}{l l c c c c c c c c c c c c c c c c c c c c c c  c  c c c c c c c c c c}
\hline
\multicolumn{1} {c}{$n$} && \multicolumn{1} {c}{$m$} &&  \multicolumn{14} {c}{$\omega^{g}$}  &&  \multicolumn{14} {c}{$\gamma$}  \\ \cline{5-18} \cline{20-33}
 && &&\multicolumn{4} {c}{$\text{Bias}^2  \times 10^2$} &&  \multicolumn{4} {c}{$\text{Variance}\times 10^2$} &&  \multicolumn{4} {c}{$\text{MSE}\times 10^2$} &&  \multicolumn{4} {c}{$\text{Bias}^2  \times 10^2$} &&  \multicolumn{4} {c}{$\text{Variance}\times 10^2$} &&  \multicolumn{4} {c}{$\text{MSE}\times 10^2$}  \\ \cline{5-8} \cline{10-13} \cline{15-18} \cline{20-23} \cline{25-28} \cline{30-33}
 	 &&		    &&	\text{OLS} & \text{Hub} & \text{Adj} & \text{QMLE} && \text{OLS} & \text{Hub} & \text{Adj} & \text{QMLE} && \text{OLS} & \text{Hub} & \text{Adj} & \text{QMLE} && \text{OLS} & \text{Hub} & \text{Adj} & \text{QMLE} && \text{OLS} & \text{Hub} & \text{Adj} & \text{QMLE} && \text{OLS} & \text{Hub} & \text{Adj} & \text{QMLE} \\ \hline	
125 	 &&	390	   && 0.347 & 0.836  & 0.208 & 0.322    && 3.748 & 2.417 & 3.355 & 3.182    && 4.090 & 3.250 & 3.558 & 3.505    && 1.175 & 1.359 & 1.181 & 1.158    && 1.215 & 1.016 & 1.133 & 1.102    && 2.388 & 2.373 & 2.313 & 2.258 \\
	 &&	780	   && 0.112 & 0.427 & 0.059	 & 0.138    && 3.502 & 2.178 & 3.135 & 3.106    && 3.609 & 2.602 & 3.189 & 3.240    && 1.025 & 1.200 & 1.051 & 0.951    && 1.266 & 1.070 & 1.208 & 1.267    && 2.290 & 2.269 & 2.257 & 2.217 \\
	 &&	2340    && 0.011 & 0.162 & 0.001 & 0.021    && 3.253 & 1.940 & 2.985 & 2.909    && 3.259 & 2.099 & 2.982 & 2.925    && 0.672 & 0.840 & 0.736 & 0.582    && 1.559 & 1.265 & 1.424 & 1.541    && 2.229 & 2.103 & 2.158 & 2.121 \\
	 &&	23400   && 0.003 & 0.046 & 0.020 & 0.000    && 3.162 & 1.749 & 2.940 & 2.942    && 3.161 & 1.792 & 2.955 & 2.937    && 0.297 & 0.354 & 0.322 & 0.195    && 1.837 & 1.624 & 1.804 & 1.957    && 2.131 & 1.976 & 2.123 & 2.150 \\
	 &&		   &&				&&		&& 					\\

250	  &&	390	   && 0.562 & 0.885  & 0.463 & 0.530    && 1.641 & 1.090 & 1.697 & 1.537    && 2.200 & 1.973 & 2.158 & 2.065    && 1.753 & 1.905 & 1.727 & 1.592    && 0.573 & 0.477 & 0.593 & 0.657    && 2.325 & 2.383 & 2.320 & 2.248 \\
	 &&	780	   && 0.277 & 0.501 & 0.214	 & 0.255    && 1.542 & 1.013 & 1.582 & 1.472    && 1.816 & 1.513 & 1.793 & 1.725    && 1.571 & 1.731 & 1.553 & 1.381    && 0.679 & 0.556 & 0.693 & 0.763    && 2.250 & 2.287 & 2.246 & 2.144 \\
	 &&	2340    && 0.102 & 0.225 & 0.062 & 0.081    && 1.554 & 0.953 & 1.507 & 1.442    && 1.654 & 1.176 & 1.566 & 1.521    && 1.211 & 1.340 & 1.198 & 0.976    && 0.845 & 0.700 & 0.861 & 0.980    && 2.055 & 2.039 & 2.058 & 1.955 \\
	 &&	23400   && 0.010 & 0.077 & 0.002 & 0.008    && 1.653 & 0.884 & 1.453 & 1.536    && 1.661 & 0.960 & 1.453 & 1.542    && 0.493 & 0.602 & 0.489 & 0.339    && 1.293 & 1.022 & 1.295 & 1.360    && 1.784 & 1.623 & 1.782 & 1.697 \\
	 &&		    &&				&&		&&				\\

500 	  &&	390	   && 0.713 & 0.896  & 0.602 & 0.641    && 0.960 & 0.596 & 0.909 & 0.798    && 1.671 & 1.491 & 1.510 & 1.438    && 2.005 & 2.162 & 1.976 & 1.889    && 0.354 & 0.261 & 0.374 & 0.344    && 2.360 & 2.424 & 2.351 & 2.232 \\
	 &&	780	   && 0.386 & 0.525 & 0.306	 & 0.330    && 0.857 & 0.545 & 0.879 & 0.778    && 1.242 & 1.069 & 1.184 & 1.107    && 1.831 & 1.993 & 1.796 & 1.668    && 0.397 & 0.289 & 0.426 & 0.410    && 2.227 & 2.282 & 2.220 & 2.078 \\
	 &&	2340    && 0.157 & 0.249 & 0.114 & 0.129    && 0.764 & 0.461 & 0.812 & 0.725    && 0.921 & 0.710 & 0.925 & 0.853    && 1.428 & 1.602 & 1.398 & 1.236    && 0.525 & 0.379 & 0.541 & 0.560    && 1.952 & 1.981 & 1.939 & 1.796 \\
	 &&	23400   && 0.028 & 0.074 & 0.013 & 0.022    && 0.864 & 0.568 & 0.925 & 0.836    && 0.890 & 0.642 & 0.937 & 0.857    && 0.621 & 0.741 & 0.604 & 0.458    && 0.881 & 0.621 & 0.887 & 0.929    && 1.500 & 1.362 & 1.490 & 1.385 \\ \cline{5-33}
&&  &&  \multicolumn{14} {c}{$\alpha^{g}$}  &&  \multicolumn{14} {c}{$\beta^{g}$}  \\ \cline{5-18} \cline{20-33}
 && &&\multicolumn{4} {c}{$\text{Bias}^2  \times 10^2$} &&  \multicolumn{4} {c}{$\text{Variance}\times 10^2$} &&  \multicolumn{4} {c}{$\text{MSE}\times 10^2$} &&  \multicolumn{4} {c}{$\text{Bias}^2  \times 10^2$} &&  \multicolumn{4} {c}{$\text{Variance}\times 10^2$} &&  \multicolumn{4} {c}{$\text{MSE}\times 10^2$}  \\ \cline{5-8} \cline{10-13} \cline{15-18} \cline{20-23} \cline{25-28} \cline{30-33}
 	 &&		    &&	\text{OLS} & \text{Hub} & \text{Adj} & \text{QMLE} && \text{OLS} & \text{Hub} & \text{Adj} & \text{QMLE} && \text{OLS} & \text{Hub} & \text{Adj} & \text{QMLE} && \text{OLS} & \text{Hub} & \text{Adj} & \text{QMLE} && \text{OLS} & \text{Hub} & \text{Adj} & \text{QMLE} && \text{OLS} & \text{Hub} & \text{Adj} & \text{QMLE} \\ \cline{5-33}	
125 	  &&	390	   && 0.025 & 0.006  & 0.027 & 0.035    && 0.335 & 0.238 & 0.340 & 0.291    && 0.360 & 0.244 & 0.367 & 0.327    && 1.403 & 1.530 & 1.448 & 1.351    && 0.454 & 0.379 & 0.425 & 0.462    && 1.857 & 1.909 & 1.874 & 1.813 \\
	 &&	780	   && 0.041 & 0.020 & 0.043	 & 0.049    && 0.340 & 0.237 & 0.339 & 0.312    && 0.381 & 0.257 & 0.382 & 0.361    && 1.193 & 1.319 & 1.220 & 1.093    && 0.584 & 0.469 & 0.553 & 0.631    && 1.777 & 1.788 & 1.773 & 1.723 \\
	 &&	2340    && 0.037 & 0.026 & 0.046 & 0.048    && 0.377 & 0.239 & 0.365 & 0.341    && 0.414 & 0.265 & 0.410 & 0.390    && 0.786 & 0.881 & 0.813 & 0.665    && 0.900 & 0.745 & 0.846 & 0.941    && 1.685 & 1.626 & 1.658 & 1.605 \\
	 &&	23400   && 0.017 & 0.014 & 0.021 & 0.019    && 0.432 & 0.259 & 0.408 & 0.387    && 0.449 & 0.273 & 0.428 & 0.405    && 0.244 & 0.256 & 0.250 & 0.141    && 1.420 & 1.315 & 1.384 & 1.509    && 1.662 & 1.569 & 1.633 & 1.648 \\
	 &&		    &&				&&		&& 					\\

250	&&	390	   && 0.029 & 0.012  & 0.031 & 0.039    && 0.184 & 0.124 & 0.186 & 0.164    && 0.213 & 0.137 & 0.217 & 0.203    && 1.759 & 1.853 & 1.763 & 1.636    && 0.178 & 0.140 & 0.179 & 0.231    && 1.937 & 1.993 & 1.943 & 1.867 \\
	 &&	780	   && 0.047 & 0.029 & 0.049	 & 0.057    && 0.189 & 0.121 & 0.190 & 0.174    && 0.237 & 0.149 & 0.239 & 0.231    && 1.540 & 1.657 & 1.545 & 1.395    && 0.279 & 0.216 & 0.280 & 0.344    && 1.819 & 1.873 & 1.825 & 1.739 \\
	 &&	2340    && 0.051 & 0.039 & 0.055 & 0.059    && 0.198 & 0.120 & 0.196 & 0.186    && 0.250 & 0.160 & 0.251 & 0.246    && 1.155 & 1.269 & 1.156 & 0.981    && 0.468 & 0.365 & 0.481 & 0.568    && 1.623 & 1.634 & 1.636 & 1.549 \\
	 &&	23400   && 0.023 & 0.022 & 0.028 & 0.023    && 0.248 & 0.131 & 0.233 & 0.227    && 0.271 & 0.153 & 0.260 & 0.250    && 0.333 & 0.384 & 0.337 & 0.206    && 1.028 & 0.890 & 1.020 & 1.113    && 1.360 & 1.273 & 1.356 & 1.317 \\
	 &&		   &&				&&		&&				\\

500 	 &&	390	   && 0.026 & 0.020  & 0.027 & 0.035    && 0.107 & 0.071 & 0.101 & 0.089    && 0.133 & 0.091 & 0.128 & 0.124    && 1.878 & 1.980 & 1.880 & 1.828    && 0.113 & 0.067 & 0.114 & 0.134    && 1.991 & 2.048 & 1.994 & 1.962 \\
	 &&	780	   && 0.043 & 0.037 & 0.044	 & 0.053    && 0.102 & 0.069 & 0.100 & 0.095    && 0.145 & 0.106 & 0.144 & 0.148    && 1.699 & 1.816 & 1.698 & 1.615    && 0.175 & 0.111 & 0.183 & 0.210    && 1.874 & 1.927 & 1.881 & 1.825 \\
	 &&	2340    && 0.050 & 0.048 & 0.051 & 0.057    && 0.102 & 0.066 & 0.103 & 0.104    && 0.152 & 0.115 & 0.155 & 0.161    && 1.327 & 1.468 & 1.326 & 1.177    && 0.349 & 0.225 & 0.344 & 0.370    && 1.675 & 1.693 & 1.670 & 1.547 \\
	 &&	23400   && 0.025 & 0.027 & 0.027 & 0.027    && 0.138 & 0.085 & 0.135 & 0.139    && 0.163 & 0.112 & 0.162 & 0.166    && 0.382 & 0.471 & 0.383 & 0.266    && 0.853 & 0.643 & 0.838 & 0.893    && 1.234 & 1.113 & 1.220 & 1.157 \\ \hline	 
\end{tabular}
}
\end{sidewaystable}

Tables \ref{Table-1} reports the squared biases, variances, and mean squared errors (MSEs) for the OLS, Huber, Adjusted-Huber, and QMLE methods of estimating $\theta_0$ with $n=125, 250, 500$ and $m=390, 780, 2340, 23400$. 
From Table \ref{Table-1}, we find that the MSEs usually decrease as $n$ or $m$ increases.
Also, the A-Hub estimator shows the best performance in terms of the MSE.
This may be because only A-Huber estimator can fully account for the heavy-tailed distributions.
For the bias, the OLS, Adjusted-Huber, and QMLE estimators show better performance than the Huber estimator.
This is because the Huber is the biased estimation method, while the OLS, Adjusted-Huber, and QMLE are unbiased estimation procedures. 
With the biased procedure, the Huber can obtain smaller variance, and thus, for the variance, the Huber shows the best performance. 
The Adjusted-Huber method can obtain the unbiased property, but its variance become larger than the Huber method. 
However, the Adjusted-Huber has a smaller MSE than the OLS overall. 
Furthermore, since  the QMLE method is robust to the heavy-tailed distribution, the QMLE estimator shows better performance than the OLS, but the QMLE does not perform better than the Huber. 
This may be because the QMLE does not account for the heavy-tailed observation directly. 
These results support the theoretical results in Section \ref{SEC-3}.

\begin{figure}[!ht]
\centering
\includegraphics[width = 1\textwidth]{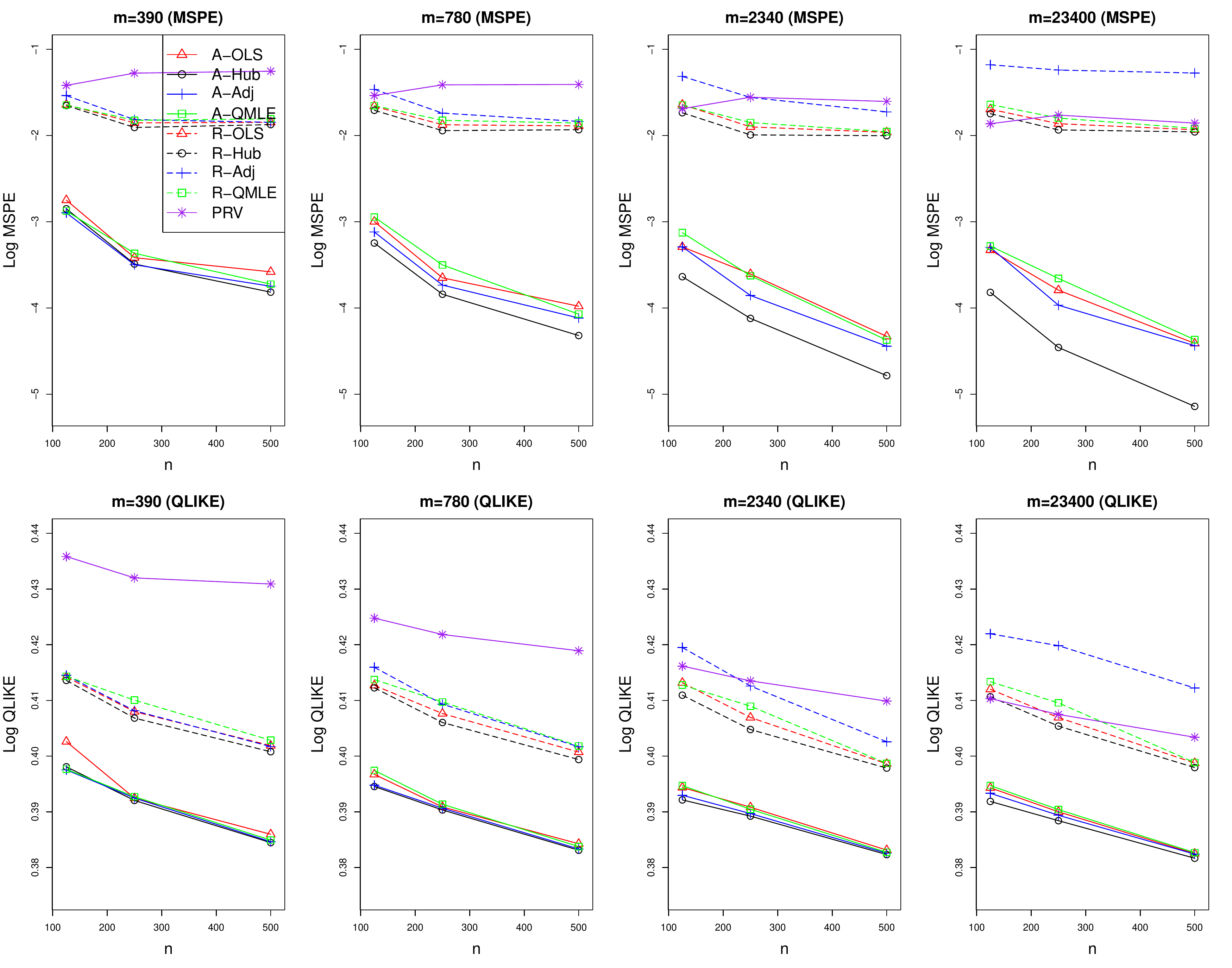}
\caption{The log MSPEs and QLIKEs for the A-OLS, A-Hub, A-Adj, A-QMLE, R-OLS, R-Hub, R-Adj, R-QMLE, and PRV of predicting $h_{n+1}(\theta_{0} )$ for $n=125, 250, 500$ and $m=390, 780, 2340, 23400$.}\label{Fig-1}
\end{figure}

To check the prediction performance, we examined the out-of-sample performance of estimating the one-day-ahead GARCH volatility $h_{n+1}\left(\theta_0 \right)$.
The OLS, Huber, Adjusted-Huber, and QMLE estimators were used to estimate future GARCH volatility based on the  ARGI  model.
We call them the A-OLS, A-Hub, A-Adj, and A-QMLE, respectively.
Also, the realized GARCH-It\^o (RGI) model is employed with the OLS, Huber, Adjusted-Huber, and QMLE methods. 
Specifically, the RGI model is obtained by setting $\alpha=0$ in the ARGI model \eqref{ARGI}. 
We call them the R-OLS, R-Hub, R-Adj, and R-QMLE, respectively.
As a benchmark, we also examined the prediction performance of the previous day's pre-averaging realized volatility estimator $\hat{V}_{n}$.
We calculated the mean squared prediction error (MSPE) and QLIKE \citep{patton2011volatility} over 1000 sample paths as follows:
\begin{eqnarray*}
 && \text{MSPE} = \frac{1}{1000}\sum^{1000}_{i=1}\left(\hat{\text{var}}_{n+1,i}-h_{n+1,i}(\theta_0 )   \right)^2, \cr
 && \text{QLIKE} = \frac{1}{1000}\sum^{1000}_{i=1}\log \hat{\text{var}}_{n+1,i} + \dfrac{h_{n+1,i}(\theta_0 )}{\hat{\text{var}}_{n+1,i}},
\end{eqnarray*}
where $\hat{\text{var}}_{n+1,i}$ is one of the above future volatility estimators for the $i$-th sample path with the information available at time $n$. 
Figure \ref{Fig-1} plots the MSPE and QLIKE for each future volatility estimator with $n=125, 250, 500$ and $m=390, 780, 2340, 23400$.
Figure \ref{Fig-1} shows that for the ARGI model-based estimators, the MSPE and QLIKE usually decrease as $n$ or $m$ increases.
Also, the A-OLS, A-Hub, A-Adj, and A-QMLE estimators perform better than the R-OLS, R-Hub, R-Adj, and R-QMLE estimators.
This is because the RGI-based estimators do not account for the leverage effect.
On the other hand, the benchmark estimator does not work well since the rich dynamics of the ARGI model cannot be fully explained by the previous day's PRV.
When comparing estimators based on the ARGI model, the A-Huber shows the best performance. 
This may be because the A-Huber can enjoy the benefit of managing the heavy-tailed observations.
The A-Adj estimator performs slightly better than the A-OLS and A-QMLE estimators, but the improvement is less compared to that of the A-Hub estimator. 
This may be because the increased variance from the bias adjustment may cause  relatively large prediction errors.

\section{An empirical study}  \label{SEC-5}
In this section, we applied the proposed ARGI model to high-frequency trading data from January $2012$ to December $2017$ (1509 days). The intra-day data for the S\&P 500 index were collected from the Wharton Research Data Services (WRDS) system. 
We defined one day as the open-to-open period. 
We employed \eqref{PRV-total}--\eqref{PRV-integratedvol} to estimate open-to-close total variation and integrated volatility. 
We chose $c_{\varpi}$ as 7 times the sample standard deviation for the pre-averaged variables $m^{1/8}\bar{Y}\left(t_{d,k}\right)$. For the close-to-open period, we considered the squared log-return as the integrated volatility. 
We applied the rolling window scheme with the in-sample period of 125 days. 
We used three different out-of-sample periods: day 126 to day 1509 (period 1), day 126 to day 817 (period 2), and day 818 to day 1509 (period 3).
We note that for the last sample period, the A-Hub estimates are $\omega^{g}=5.883 \times 10^{-6}$, $\gamma=2.474 \times 10^{-1}$, $\alpha^{g}=6.937 \times 10^{-4}$, and $\beta^{g}=2.225 \times 10^{-1}$.

\begin{figure}[!ht]
\centering
\includegraphics[width = 0.9\textwidth]{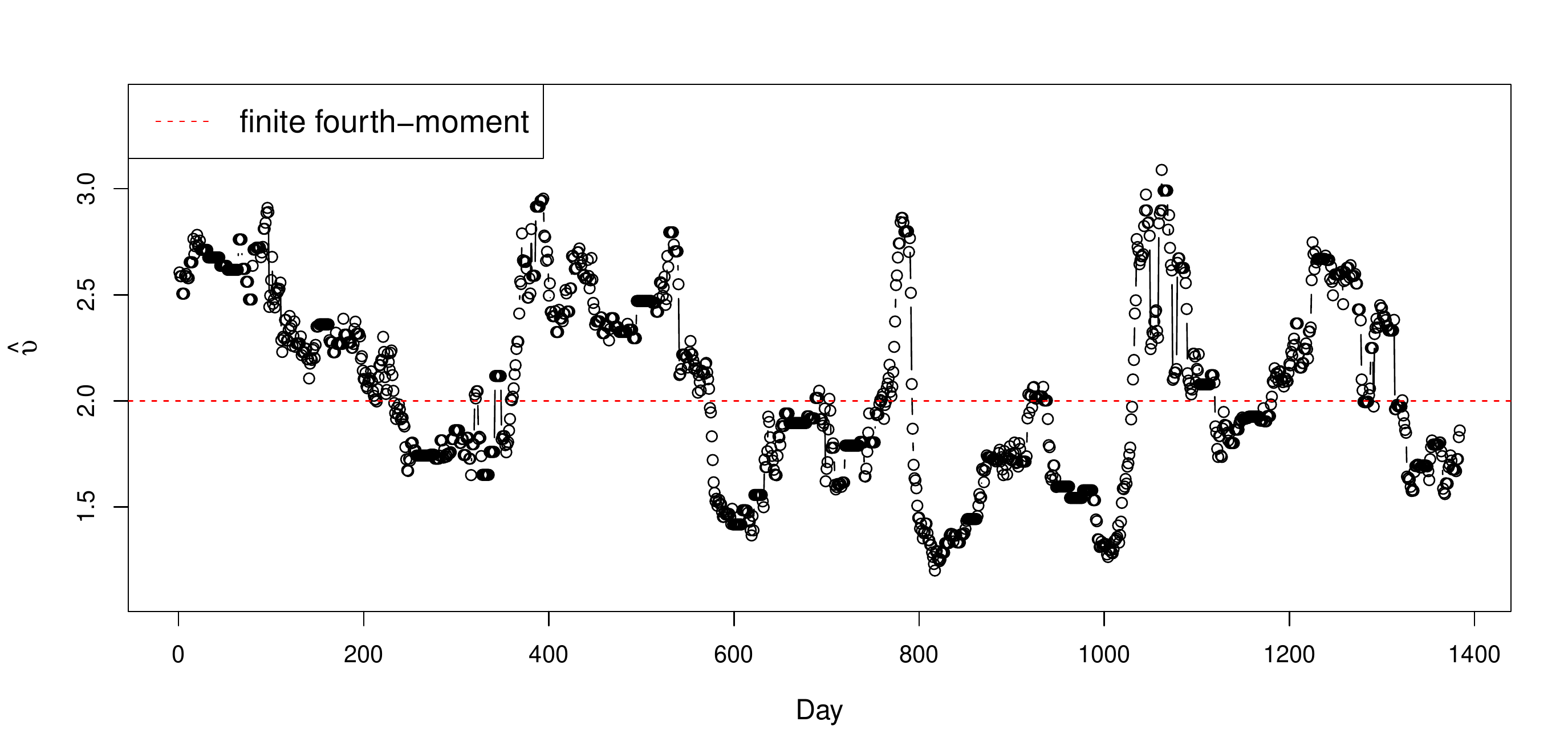}
\caption{Daily estimated tail indices $\hat{\upsilon}$ obtained from the Hill's estimator. The red dash line represents the existence of finite fourth-moment for the log-returns of S\&P 500 index.}\label{Fig-2}
\end{figure}

To catch the heavy-tailedness in financial data, we estimated the tail indices using the Hill's estimator $\hat{\upsilon}$ in  \eqref{Hill}. 
Figure \ref{Fig-2} draws the daily estimated tail index $\hat{\upsilon}$ for  each in-sample data.
From Figure \ref{Fig-2}, we find that the financial data often exhibit
severe heavy-tailedness. This supports the heavy-tailedness condition Assumption \ref{assumption1}(b).

To investigate the performance of the proposed estimation methods, we calculated the one-day ahead conditional expected volatility estimator $\text{Vol}_{d}$ based on the A-OLS, A-Hub, A-Adj, A-QMLE, R-OLS, R-Hub, R-Adj, and R-QMLE estimators.
For a comparison, we employed the heterogeneous auto-regressive (HAR) model \citep{corsi2009simple}.
Specifically, we first implemented  the following linear regression:
\begin{equation*}
 RV_{d+1}=a+b_{1} RV_{d} + b_{2} RV_{d}^{(week)} + b_{3} RV_{d}^{(month)}+e_{d+1},
\end{equation*}
where $RV_{d}$ is defined in \eqref{PRV-integratedvol},
\begin{equation*}
 RV_{d}^{(week)}=\frac{1}{5}\sum^{4}_{i=0}RV_{d-i}, \quad \text{ and } \quad RV_{d}^{(month)}=\frac{1}{22}\sum^{21}_{i=0}RV_{d-i}.
\end{equation*}
We then obtained the conditional expected volatility for the next day and call it the HAR estimator.
We note that for the HAR model, due to its high sensitivity for one outlier point, we bounded the largest prediction value to the second-largest value, which helps improve the performance of the HAR model.
To avoid the negative value for the volatility estimates, we set the lower bound of $\text{Vol}_{d}$ as $10^{-5}$.

With the volatility estimator $\text{Vol}_{d}$,  we employed the mean squared prediction error (MSPE), relative mean squared prediction error (RMSPE), and QLIKE \citep{patton2011volatility} as follows:
\begin{eqnarray*}
&& \text{MSPE}=\frac{1}{n}\sum^{n}_{d=1}\left(\text{Vol}_{d}-\hat{V}_{d}\right)^2, \quad  \text{RMSPE}=\frac{1}{n}\sum^{n}_{d=1}\left(\frac{\text{Vol}_{d}-\hat{V}_{d}}{\hat{V}_{d}}\right)^2, \cr
&& \text{QLIKE}=\frac{1}{n}\sum^{n}_{d=1}\log \text{Vol}_{d} + \dfrac{\hat{V}_{d}}{\text{Vol}_{d}},
\end{eqnarray*}
where $\hat{V}_{d}$ is the non-parametric estimator of daily total variation defined in \eqref{PRV-total}.
Also, we conducted the Diebold and Mariano (DM) test \citep{diebold1995paring} to investigate the significance of the difference in performance.
Specifically, the DM test checks whether the expectation of the loss differential series, $Z_d = f(\text{Vol}_{1,d},\hat{V}_{d}) - f(\text{Vol}_{2,d},\hat{V}_{d})$, is zero for estimators $j=1,2$.
Under the null hypothesis $\E(Z_d)=0$, the test statistic $DM = \bar{Z}/\hat{\sigma}_{Z} \overset{d}{\to} N(0,1)$, where $\bar{Z}$ is the sample mean of $Z_d$ and $\hat{\sigma}_Z$ is a consistent estimator of the standard deviation. 
For the loss function $f(x,y)$, we used $(x-y)^2$, $\log x + \dfrac{y}{x}$, and $\left(\dfrac{x-y}{y}\right)^2$ for the MSPE, RMSPE, and QLIKE, respectively. 
For the MSPE and RMSPE, we  set $j=1$ for the A-Hub estimator and $j=2$ for other estimators, while for QLIKE, we set $j=1$ for the A-QMLE and $j=2$ for other estimators.

\begin{table}[!ht]
\caption{The MSPE, RMSPE, and QLIKE of the A-OLS, A-Hub, A-Adj, A-QMLE, R-OLS, R-Hub, R-Adj, R-QMLE, and HAR estimators
for three out-of-sample periods.}\label{Table-2}
\centering
\scalebox{0.67}{
\begin{tabular}{l l l l c c c c c c c c c c c c c c c c c}
\hline
  && && \text{A-OLS} && \text{A-Hub} && \text{A-Adj}  && \text{A-QMLE} && \text{R-OLS} && \text{R-Hub} && \text{R-Adj} && \text{R-QMLE} && \text{HAR} \\ \hline
Period 1 && $\text{MSPE}\times 10^9$ 	 &&	1.552     &&	1.319  	&&  1.548  	&&	 1.338  &&	1.379 	&&  1.377  &&  1.436	 &&  1.343	&&  1.426		\\
&&$\text{RMSPE}$ 	 &&	0.930     &&	0.494  	&& 1.101 	&&	 0.870  &&	1.022 	&&  0.541  	&&  1.128  &&  0.937	 &&  0.572		\\ 
&&$\text{QLIKE}$ 	 &&	-9.535     &&	-9.523  	&& -9.472 	&&	 -9.547  &&	 -9.541 	&&  -9.517  	&&  -9.538  &&  -9.543	 &&  -9.493		\\ 
&&                 &&           &&            &&       &&          &&         &&                     \\

Period 2 && $\text{MSPE}\times 10^9$ 	 &&	0.498     &&	0.477  	&& 0.659  	&&	 0.488  &&	0.506 	&&  0.495  &&  0.653	 &&  0.494 &&  0.553	\\
&&$\text{RMSPE}$ 	 &&	0.856     &&	0.452  	&&  1.318 	&&	 0.849  &&	0.955 	&&  0.500  	&&  1.131  	&&  0.905 &&  0.592 \\ 
&&$\text{QLIKE}$ 	 &&	-9.505     &&	-9.487  	&& -9.453 	&&	 -9.509  &&	-9.503 	&&  -9.483  	&&  -9.501  	&&  -9.503 &&  -9.468 \\ 
&&                 &&           &&            &&       &&          &&         &&                      \\

Period 3 && $\text{MSPE}\times 10^9$ 	 &&	2.606     &&	2.162  	&& 2.438  	&&	 2.189  &&	2.252 	&&  2.259  &&  2.219	 &&  2.192 &&  2.298	\\
&&$\text{RMSPE}$ 	 &&	1.004     &&	0.536  	&& 0.885 	&&	 0.892  &&	1.089 	&&  0.583  	&&  1.124  	&&  0.969 &&  0.553 \\ 
&&$\text{QLIKE}$ 	 &&	-9.565     &&	-9.560  	&& -9.492 	&&	 -9.584  &&	 -9.579 	&&  -9.552  	&&  -9.574  &&  -9.583 &&  -9.518 \\   \hline
\end{tabular}
}
\end{table}

\begin{table}[!ht]
\caption{The test statistic DM based on the MSPE and RMSPE for the A-OLS, A-Adj, A-QMLE, R-OLS, R-Hub, R-Adj, R-QMLE, and HAR estimators with respect to the A-Hub estimator for three out-of-sample periods. In the parentheses, we report the corresponding p-values.}\label{Table-3}
\centering
\scalebox{0.62}{
\begin{tabular}{l l l l c c c c c cc c}
\hline
 &&  && \text{A-OLS} & \text{A-Adj} &  \text{A-QMLE} & \text{R-OLS} & \text{R-Hub} & \text{R-Adj}  & \text{R-QMLE} & \text{HAR} \\ \hline
Period 1 &&	 $\text{MSPE}$   &&	-0.71 (0.23) 	& -2.86 (0.00)    &  -0.27 (0.39)	&	 -0.48 (0.31) &	 -2.42 (0.00) 	&  -1.08 (0.13) 	&  -0.24 (0.40) 	&  -1.69 (0.04)  	\\
 &&	 $\text{RMSPE}$   &&	-10.65 (0.00) 	& -7.94 (0.00)    &  -9.05 (0.00)	&	 -11.09 (0.00) &	 -3.95 (0.00) 	&  -6.45 (0.00) 	&  -9.93 (0.00) 	&  -2.90 (0.00)  	\\
&& && \\

Period 2 &&	 $\text{MSPE}$   &&	-0.98 (0.16) 	& -2.12 (0.01)   &  -0.52 (0.30)	&  -1.11 (0.13) &	 -2.31 (0.01) 	&  -1.64 (0.05) 	&  -1.17 (0.12) 	&  -3.23 (0.00)  	\\
 &&	 $\text{RMSPE}$   &&	-9.44 (0.00) 	& -8.39 (0.00)    &  -8.04 (0.00)	&	 -9.15 (0.00) &	 -3.20 (0.00) 	&  -5.63 (0.00) 	&  -8.39 (0.00) 	&  -4.07 (0.00)  	\\
&& && \\

Period 3 &&	 $\text{MSPE}$   &&	-0.97 (0.16) 	& -3.72 (0.00)   &  -0.27 (0.39)	&	 -0.52 (0.29) &	 -2.97 (0.00) 	&  -0.53 (0.29)  	&  -0.22 (0.41) 	&  -1.59 (0.05)  	\\ 
 &&	 $\text{RMSPE}$   &&	-12.01 (0.00) 	& -11.22 (0.00)    &  -11.12 (0.00)	&	 -14.23 (0.00) &	 -5.95 (0.00) 	&  -8.57 (0.00) 	&  -13.24 (0.00) 	&  -1.00 (0.15)  	\\ \hline
\end{tabular}
}
\end{table}

\begin{table}[!ht]
\caption{The test statistic DM based on the QLIKE for the A-OLS, A-Hub, A-Adj, R-OLS, R-Hub, R-Adj, R-QMLE, and HAR estimators with respect to the A-QMLE estimator for three out-of-sample periods. In the parentheses, we report the corresponding p-values.}\label{Table-4}
\centering
\scalebox{0.62}{
\begin{tabular}{l l l l c c c c c c c c}
\hline
 &&  && \text{A-OLS} & \text{A-Hub} &  \text{A-Adj} & \text{R-OLS} & \text{R-Hub} & \text{R-Adj}  & \text{R-QMLE} & \text{HAR} \\ \hline
Period 1 &&	 $\text{QLIKE}$   &&	-1.81 (0.03) 	& -3.88 (0.00)    &  -6.23 (0.00)	&	 -1.72 (0.04) &	 -4.98 (0.00) 	&  -2.61 (0.00) 	&  -0.98 (0.16) 	&  -5.46 (0.00)  	\\
&& && \\

Period 2 &&	 $\text{QLIKE}$   &&	-1.42 (0.07) 	& -4.24 (0.00)   &  -6.42 (0.00)	&  -2.67 (0.00) &	 -5.46 (0.00) 	&  -3.02 (0.00) 	&  -3.05 (0.00) 	&  -7.45 (0.00)  	\\
&& && \\

Period 3 &&	 $\text{QLIKE}$   &&	-2.22 (0.01) 	& -3.65 (0.00)   &  -6.40 (0.00)	&	 -1.24 (0.10) &	 -4.78 (0.00) 	&  -2.43 (0.00)  	&  -0.11 (0.45) 	&  -5.17 (0.00)  	\\  \hline
\end{tabular}
}
\end{table}

Table \ref{Table-2} reports the MSPE, RMSPE, and QLIKE of the  A-OLS, A-Hub, A-Adj, A-QMLE, R-OLS, R-Hub, R-Adj, R-QMLE, and HAR estimators for each out-of-sample period.
Tables \ref{Table-3}-\ref{Table-4} show the test statistics of the DM tests with the corresponding p-values.
As seen in Tables \ref{Table-2}--\ref{Table-3}, the A-Hub estimator outperforms other estimators for the MSPE and RMSPE.
This result indicates that considering both heavy-tailedness and leverage effect  helps explain the market dynamics. 
On the other hand, as seen in Tables \ref{Table-2} and \ref{Table-4}, the A-QMLE estimator shows the best performance for the QLIKE.
This may be because the QMLE estimator is obtained based on the QLIKE loss and the effect of heavy-tailed observations is relatively small for the QLIKE loss.
In contrast, the A-Adj estimator shows relatively worse performance. 
One possible explanation for this is that since  the log-returns often exhibit severe heavy-tailedness as shown in Figure \ref{Fig-2}, the loss from the increased variance is larger than the gain from the reduced bias when predicting volatilities with the bias adjustment scheme.

\begin{figure}[!ht]
\centering
\includegraphics[width = 1\textwidth]{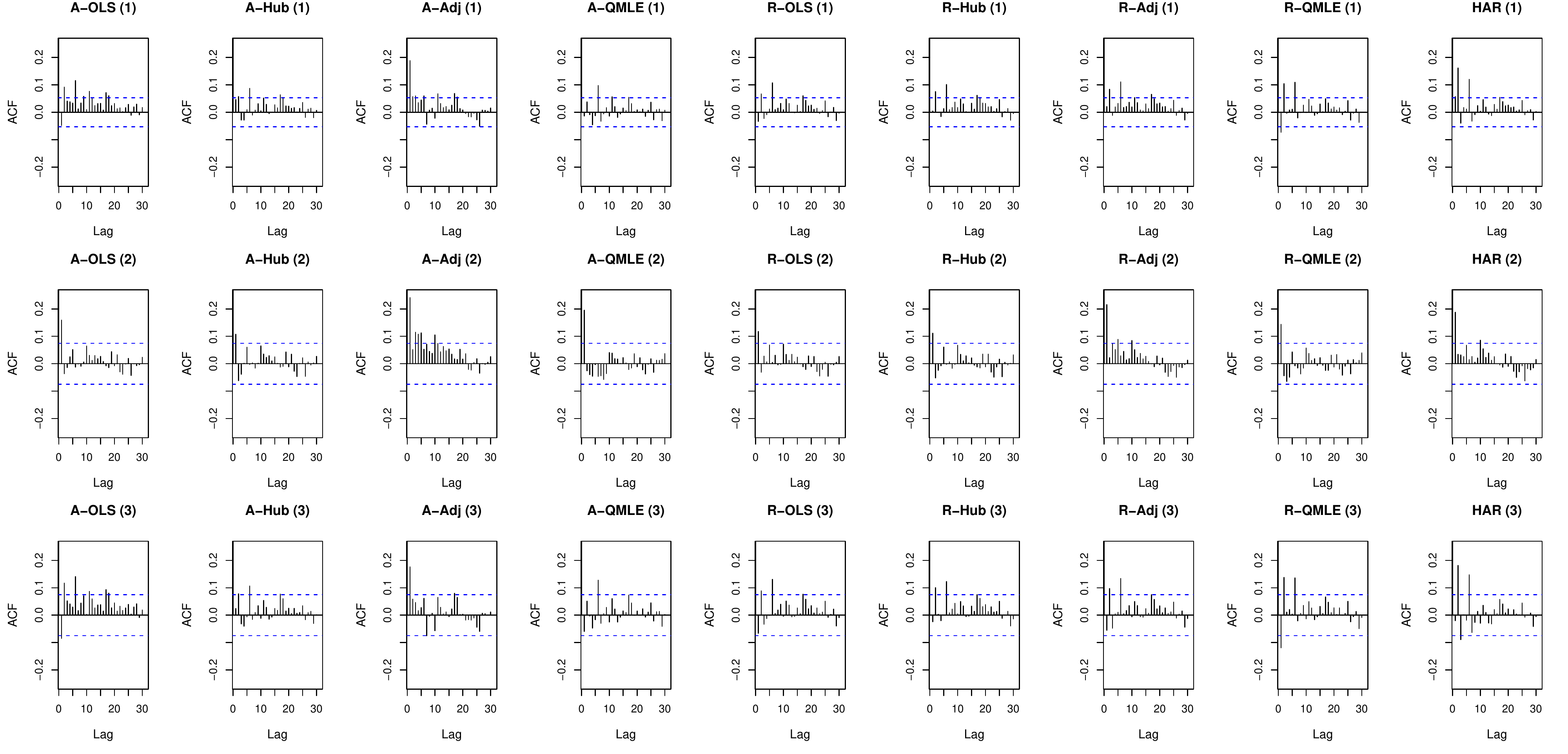}
\caption{The ACF plots of the regression residuals obtained from the future volatility estimators A-OLS, A-Hub, A-Adj, A-QMLE, R-OLS, R-Hub, R-Adj, R-QMLE, and HAR for three out-of-sample periods. Note that (1), (2), and (3) represent the periods 1, 2, and 3, respectively.}\label{Fig-3}
\end{figure}

We also tested the proposed estimators using the persistence of the non-parametric volatility  \citep{brandt2006volatility, Kim2021Overnight}.
Specifically, we conducted  the following linear regression between the non-parametric volatility $\hat{V}_d$ and future volatility estimator $\text{Vol}_{d}$:
\begin{equation*}
\hat{V}_d = a + b \times \text{Vol}_{d} + e_d,
\end{equation*}
where $\text{Vol}_{d}$ is one of the future volatility estimators A-OLS, A-Hub, A-Adj, A-QMLE, R-OLS, R-Hub, R-Adj, R-QMLE, and HAR. 
We then obtained the regression residuals, $\hat{\epsilon}_d$, and calculated their auto-correlations over lag $L=1, \ldots, 30$.  
Table \ref{Table-5} reports the largest absolute auto-correlation over lag $L=1, \ldots, 30$ and Figure \ref{Fig-3} draws the ACF plot for each future volatility estimator and out-of-sample period. From Table \ref{Table-5} and Figure \ref{Fig-3}, we find that the Huber estimation method performs best for both ARGI- and RGI-based estimators.
From this result, we can conjecture that considering heavy-tailedness can help reduce the volatility persistence.
Also, we find that the A-Hub estimator shows the best performance.
It provides the evidence to conclude that the A-Hub estimation method helps account for the market dynamics.
From these results, we can conjecture that considering the stylized facts of financial data helps better account for market dynamics, and ignoring one of them often misses some market dynamics.

\begin{table}[!ht]
\caption{The largest absolute auto-correlation of the regression residuals over lag $L=1, \ldots, 30$ obtained from the A-OLS, A-Hub, A-Adj, A-QMLE, R-OLS, R-Hub, R-Adj, R-QMLE, and HAR estimators for three out-of-sample periods.}\label{Table-5}
\centering
\scalebox{0.77}{
\begin{tabular}{l l c c c c c c c c c c c c c c c c c c c}
\hline
  && \text{A-OLS} && \text{A-Hub} && \text{A-Adj} && \text{A-QMLE} && \text{R-OLS} && \text{R-Hub} && \text{R-Adj} && \text{R-QMLE} && \text{HAR}   \\ \hline
Period 1 	    &&	0.115     &&	0.087 	&&  0.188 	&&	 0.097  &&	0.107 	&&  0.101  &&  0.110 	&&  0.109  &&  0.161 	\\
Period 2	    &&	0.159     &&	0.107 	&&  0.241 	&&	 0.195  &&	0.117 	&&  0.111  &&  0.215 	&&  0.143  &&  0.187 	\\
Period 3     &&	0.140     &&	0.106 	&&  0.176 	&&	 0.127  &&	0.130 	&&  0.122  &&  0.133 	&&  0.137  &&  0.181 \\ \hline
\end{tabular}
}
\end{table}

\section{Conclusions}  \label{SEC-6}
In this paper, we develop an asymmetric realized GARCH-It\^o (ARGI) model to account for the well-known stylized facts of high-frequency financial data such as the leverage effect, volatility clustering, price jump, heavy-tailedness, and intra-day U-shape.
Under the ARGI model, we show that the conditional daily total variation has the autoregressive structure with the asymmetric effect of the daily log-return. 
To deal with the heavy-tailedness of the financial data, we propose a  Huber regression procedure.
We show that it has the asymptotic normality with the optimal convergence rate.
Also, to handle the bias coming from truncation in Huber regression and asymmetry in financial observations, we suggest a bias adjustment scheme  and obtain the unbiased estimator which has the stable limit property with the optimal rate.

In the empirical study, for the prediction of total variation and analysis of volatility persistence, the Huber estimator based on the ARGI model shows the best performance. 
It suggests that considering the leverage effect and heavy-tailedness of financial data can help account for market dynamics. 
On the other hand, the Adjusted-Huber estimator with the ARGI model shows relatively worse performance. 
This may be because the bias adjustment term also has the heavy-tailedness, which can cause estimation errors. 
Finally, from Figure \ref{Fig-2}, we find some heterogeneous degrees of heavy-tailedness over the time period.
However, the current Huber regression procedure assumes that during the in-sample period,
the heavy-tailedness is homogeneous, which may cause inefficiency of estimating parameters.
To account for the heterogeneous heavy-tailedness, we may need to introduce several tuning parameters, which may bring some severe tuning parameter choice problem.
Thus, it would be interesting and important to develop the adaptive estimation method which can handle the time-varying heavy-tailedness and solve the tuning parameter choice problem.
We leave this issue for future study.

\section*{Funding} 
 
This work was supported by the National Research Foundation of Korea [2021R1C1C1003216].

\section*{Data Availability Statement} 

The S\&P 500 intraday index data is provided by the Wharton Research Data Services (WRDS) system (web link: https://wrds-www.wharton.upenn.edu/). Please note that the data sharing policy of WRDS restricts the redistribution of data.

\bibliography{myReferences}

\begin{thebibliography}{}

\bibitem[Admati and Pfleiderer, 1988]{admati1988theory}
Admati, A.~R. and Pfleiderer, P. (1988).
\newblock A theory of intraday patterns: Volume and price variability.
\newblock {\em The Review of Financial Studies}, 1(1):3--40.

\bibitem[A{\"\i}t-Sahalia et~al., 2017]{ait2017estimation}
A{\"\i}t-Sahalia, Y., Fan, J., Laeven, R.~J., Wang, C.~D., and Yang, X. (2017).
\newblock Estimation of the continuous and discontinuous leverage effects.
\newblock {\em Journal of the American Statistical Association},
  112(520):1744--1758.

\bibitem[A{\"\i}t-Sahalia et~al., 2010]{ait2010high}
A{\"\i}t-Sahalia, Y., Fan, J., and Xiu, D. (2010).
\newblock High-frequency covariance estimates with noisy and asynchronous
  financial data.
\newblock {\em Journal of the American Statistical Association},
  105(492):1504--1517.

\bibitem[A{\"\i}t-Sahalia et~al., 2012]{ait2012testing}
A{\"\i}t-Sahalia, Y., Jacod, J., and Li, J. (2012).
\newblock Testing for jumps in noisy high frequency data.
\newblock {\em Journal of Econometrics}, 168(2):207--222.

\bibitem[A{\"\i}t-Sahalia and Xiu, 2016]{ait2016increased}
A{\"\i}t-Sahalia, Y. and Xiu, D. (2016).
\newblock Increased correlation among asset classes: Are volatility or jumps to
  blame, or both?
\newblock {\em Journal of Econometrics}, 194(2):205--219.

\bibitem[Andersen et~al., 2007]{andersen2007roughing}
Andersen, T.~G., Bollerslev, T., and Diebold, F.~X. (2007).
\newblock Roughing it up: Including jump components in the measurement,
  modeling, and forecasting of return volatility.
\newblock {\em The review of economics and statistics}, 89(4):701--720.

\bibitem[Andersen et~al., 2003]{andersen2003modeling}
Andersen, T.~G., Bollerslev, T., Diebold, F.~X., and Labys, P. (2003).
\newblock Modeling and forecasting realized volatility.
\newblock {\em Econometrica}, 71(2):579--625.

\bibitem[Andersen et~al., 1997]{andersen1997intraday}
Andersen, T.~G., Bollerslev, T., et~al. (1997).
\newblock Intraday periodicity and volatility persistence in financial markets.
\newblock {\em Journal of empirical finance}, 4(2-3):115--158.

\bibitem[Andersen et~al., 2019]{andersen2018time}
Andersen, T.~G., Thyrsgaard, M., and Todorov, V. (2019).
\newblock Time-varying periodicity in intraday volatility.
\newblock {\em Journal of the American Statistical Association}.

\bibitem[Andrews, 1992]{andrews1992generic}
Andrews, D.~W. (1992).
\newblock Generic uniform convergence.
\newblock {\em Econometric theory}, 8(2):241--257.

\bibitem[Barndorff-Nielsen et~al., 2008]{barndorff2008designing}
Barndorff-Nielsen, O.~E., Hansen, P.~R., Lunde, A., and Shephard, N. (2008).
\newblock Designing realized kernels to measure the ex post variation of equity
  prices in the presence of noise.
\newblock {\em Econometrica}, 76(6):1481--1536.

\bibitem[Barndorff-Nielsen and Shephard, 2006]{barndorff2006econometrics}
Barndorff-Nielsen, O.~E. and Shephard, N. (2006).
\newblock Econometrics of testing for jumps in financial economics using
  bipower variation.
\newblock {\em Journal of financial Econometrics}, 4(1):1--30.

\bibitem[Bauwens et~al., 2014]{bauwens2014marginal}
Bauwens, L., Dufays, A., and Rombouts, J.~V. (2014).
\newblock Marginal likelihood for markov-switching and change-point garch
  models.
\newblock {\em Journal of Econometrics}, 178:508--522.

\bibitem[Bauwens et~al., 2010]{bauwens2010theory}
Bauwens, L., Preminger, A., and Rombouts, J.~V. (2010).
\newblock Theory and inference for a markov switching garch model.
\newblock {\em Econometrics Journal}, 13(2):218--244.

\bibitem[Black, 1976]{black1976studies}
Black, F. (1976).
\newblock Studies of stock market volatility changes.
\newblock {\em 1976 Proceedings of the American Statistical Association
  Bisiness and Economic Statistics Section}.

\bibitem[Bollerslev, 1986]{bollerslev1986generalized}
Bollerslev, T. (1986).
\newblock Generalized autoregressive conditional heteroskedasticity.
\newblock {\em Journal of econometrics}, 31(3):307--327.

\bibitem[Brandt and Jones, 2006]{brandt2006volatility}
Brandt, M.~W. and Jones, C.~S. (2006).
\newblock Volatility forecasting with range-based egarch models.
\newblock {\em Journal of Business \& Economic Statistics}, 24(4):470--486.

\bibitem[Brown et~al., 1971]{brown1971martingale}
Brown, B.~M. et~al. (1971).
\newblock Martingale central limit theorems.
\newblock {\em Annals of Mathematical Statistics}, 42(1):59--66.

\bibitem[Christie, 1982]{christie1982stochastic}
Christie, A.~A. (1982).
\newblock The stochastic behavior of common stock variances: Value, leverage
  and interest rate effects.
\newblock {\em Journal of Financial Economics}, 10(4):407--432.

\bibitem[Chun and Kim, 2022]{chun2022state}
Chun, D. and Kim, D. (2022).
\newblock State heterogeneity analysis of financial volatility using
  high-frequency financial data.
\newblock {\em Journal of Time Series Analysis}, 43(1):105--124.

\bibitem[Cont, 2001]{cont2001empirical}
Cont, R. (2001).
\newblock Empirical properties of asset returns: stylized facts and statistical
  issues.
\newblock {\em Quantitative Finance}, 1(2):223--236.

\bibitem[Corsi, 2009]{corsi2009simple}
Corsi, F. (2009).
\newblock A simple approximate long-memory model of realized volatility.
\newblock {\em Journal of Financial Econometrics}, 7(2):174--196.

\bibitem[Corsi et~al., 2010]{corsi2010threshold}
Corsi, F., Pirino, D., and Reno, R. (2010).
\newblock Threshold bipower variation and the impact of jumps on volatility
  forecasting.
\newblock {\em Journal of Econometrics}, 159(2):276--288.

\bibitem[Diebold and Mariano, 1995]{diebold1995paring}
Diebold, F.~X. and Mariano, R.~S. (1995).
\newblock Comparing predictive accuracy.
\newblock {\em Journal of Business and Economic Statistics}, 13(3):253--263.

\bibitem[Engle, 1982]{engle1982autoregressive}
Engle, R.~F. (1982).
\newblock Autoregressive conditional heteroscedasticity with estimates of the
  variance of united kingdom inflation.
\newblock {\em Econometrica: Journal of the Econometric Society}, pages
  987--1007.

\bibitem[Engle and Ng, 1993]{engle1993measuring}
Engle, R.~F. and Ng, V.~K. (1993).
\newblock Measuring and testing the impact of news on volatility.
\newblock {\em The journal of finance}, 48(5):1749--1778.

\bibitem[Fan et~al., 2019]{fan2019adaptive}
Fan, J., Guo, Y., and Jiang, B. (2019).
\newblock Adaptive huber regression on markov-dependent data.
\newblock {\em Stochastic Processes and their Applications}.

\bibitem[Fan and Kim, 2018]{fan2018robust}
Fan, J. and Kim, D. (2018).
\newblock Robust high-dimensional volatility matrix estimation for
  high-frequency factor model.
\newblock {\em Journal of the American Statistical Association},
  113(523):1268--1283.

\bibitem[Fan and Wang, 2007]{fan2007multi}
Fan, J. and Wang, Y. (2007).
\newblock Multi-scale jump and volatility analysis for high-frequency financial
  data.
\newblock {\em Journal of the American Statistical Association},
  102(480):1349--1362.

\bibitem[Figlewski and Wang, 2000]{figlewski2000leverage}
Figlewski, S. and Wang, X. (2000).
\newblock Is the `leverage effect' a leverage effect?
\newblock {\em Available at SSRN 256109}.

\bibitem[Glosten et~al., 1993]{glosten1993relation}
Glosten, L.~R., Jagannathan, R., and Runkle, D.~E. (1993).
\newblock On the relation between the expected value and the volatility of the
  nominal excess return on stocks.
\newblock {\em Journal of Finance}, 48(5):1779--1801.

\bibitem[Gray, 1996]{gray1996modeling}
Gray, S.~F. (1996).
\newblock Modeling the conditional distribution of interest rates as a
  regime-switching process.
\newblock {\em Journal of Financial Economics}, 42(1):27--62.

\bibitem[Haas et~al., 2004]{haas2004new}
Haas, M., Mittnik, S., and Paolella, M.~S. (2004).
\newblock A new approach to markov-switching garch models.
\newblock {\em Journal of Financial Econometrics}, 2(4):493--530.

\bibitem[Hall and Heyde, 2014]{hall2014martingale}
Hall, P. and Heyde, C.~C. (2014).
\newblock {\em Martingale limit theory and its application}.
\newblock Academic press.

\bibitem[Hamilton and Susmel, 1994]{hamilton1994autoregressive}
Hamilton, J.~D. and Susmel, R. (1994).
\newblock Autoregressive conditional heteroskedasticity and changes in regime.
\newblock {\em Journal of Econometrics}, 64(1-2):307--333.

\bibitem[Hansen et~al., 2012]{hansen2012realized}
Hansen, P.~R., Huang, Z., and Shek, H.~H. (2012).
\newblock Realized garch: a joint model for returns and realized measures of
  volatility.
\newblock {\em Journal of Applied Econometrics}, 27(6):877--906.

\bibitem[Hill, 1975]{hill1975simple}
Hill, B.~M. (1975).
\newblock A simple general approach to inference about the tail of a
  distribution.
\newblock {\em The annals of statistics}, pages 1163--1174.

\bibitem[Hill, 2010]{hill2010tail}
Hill, J.~B. (2010).
\newblock On tail index estimation for dependent, heterogeneous data.
\newblock {\em Econometric Theory}, pages 1398--1436.

\bibitem[Hong and Wang, 2000]{hong2000trading}
Hong, H. and Wang, J. (2000).
\newblock Trading and returns under periodic market closures.
\newblock {\em The Journal of Finance}, 55(1):297--354.

\bibitem[Hsing, 1991]{hsing1991tail}
Hsing, T. (1991).
\newblock On tail index estimation using dependent data.
\newblock {\em The Annals of Statistics}, pages 1547--1569.

\bibitem[Jacod et~al., 2009]{jacod2009microstructure}
Jacod, J., Li, Y., Mykland, P.~A., Podolskij, M., and Vetter, M. (2009).
\newblock Microstructure noise in the continuous case: the pre-averaging
  approach.
\newblock {\em Stochastic Processes and their Applications}, 119(7):2249--2276.

\bibitem[Jiang and Yao, 2013]{jiang2013stock}
Jiang, G.~J. and Yao, T. (2013).
\newblock Stock price jumps and cross-sectional return predictability.
\newblock {\em Journal of Financial and Quantitative Analysis},
  48(5):1519--1544.

\bibitem[Kim, 2016]{kim2016statistical}
Kim, D. (2016).
\newblock Statistical inference for unified garch--it{\^o} models with
  high-frequency financial data.
\newblock {\em Journal of Time Series Analysis}, 37(513--532).

\bibitem[Kim and Fan, 2019]{kim2019factor}
Kim, D. and Fan, J. (2019).
\newblock Factor garch-it{\^o} models for high-frequency data with application
  to large volatility matrix prediction.
\newblock {\em Journal of econometrics}, 208(2):395--417.

\bibitem[Kim and Wang, 2016a]{Kim2016SPCA}
Kim, D. and Wang, Y. (2016a).
\newblock Sparse pca based on high-dimensional it\^o processes with measurement
  errors.
\newblock {\em Journal of Multivariate Analysis}, 152:172--18.

\bibitem[Kim and Wang, 2016b]{kim2016unified}
Kim, D. and Wang, Y. (2016b).
\newblock Unified discrete-time and continuous-time models and statistical
  inferences for merged low-frequency and high-frequency financial data.
\newblock {\em Journal of Econometrics}, 194:220--230.

\bibitem[Kim and Wang, 2021]{Kim2021Overnight}
Kim, D. and Wang, Y. (2021).
\newblock Overnight garch-it{\^o} volatility models.
\newblock {\em Available at SSRN 3792523}.

\bibitem[Klaassen, 2002]{klaassen2002improving}
Klaassen, F. (2002).
\newblock Improving garch volatility forecasts with regime-switching garch.
\newblock {\em Empirical Economics}, 27:363--394.

\bibitem[Mandelbrot, 1963]{mandelbrot1963variation}
Mandelbrot, B. (1963).
\newblock The variation of certain speculative prices.
\newblock {\em The journal of business}, 36(4):394--419.

\bibitem[Mao and Zhang, 2018]{mao2018stochastic}
Mao, G. and Zhang, Z. (2018).
\newblock Stochastic tail index model for high frequency financial data with
  bayesian analysis.
\newblock {\em Journal of Econometrics}, 205(2):470--487.

\bibitem[Massacci, 2017]{massacci2017tail}
Massacci, D. (2017).
\newblock Tail risk dynamics in stock returns: Links to the macroeconomy and
  global markets connectedness.
\newblock {\em Management Science}, 63(9):3072--3089.

\bibitem[Mikosch et~al., 2006]{mikosch2006stable}
Mikosch, T., Straumann, D., et~al. (2006).
\newblock Stable limits of martingale transforms with application to the
  estimation of garch parameters.
\newblock {\em The Annals of Statistics}, 34(1):493--522.

\bibitem[Nyberg, 2012]{nyberg2012risk}
Nyberg, H. (2012).
\newblock Risk-return tradeoff in us stock returns over the business cycle.
\newblock {\em Journal of Financial and Quantitative Analysis}, 47(1):137--158.

\bibitem[Patton, 2011]{patton2011volatility}
Patton, A.~J. (2011).
\newblock Volatility forecast comparison using imperfect volatility proxies.
\newblock {\em Journal of Econometrics}, 160(1):246--256.

\bibitem[Shephard and Sheppard, 2010]{shephard2010realising}
Shephard, N. and Sheppard, K. (2010).
\newblock Realising the future: forecasting with high-frequency-based
  volatility (heavy) models.
\newblock {\em Journal of Applied Econometrics}, 25(2):197--231.

\bibitem[Shin et~al., 2021]{shin2021adaptive}
Shin, M., Kim, D., and Fan, J. (2021).
\newblock Adaptive robust large volatility matrix estimation based on
  high-frequency financial data.
\newblock {\em arXiv preprint arXiv:2102.12752}.

\bibitem[Song et~al., 2021]{song2021volatility}
Song, X., Kim, D., Yuan, H., Cui, X., Lu, Z., Zhou, Y., and Wang, Y. (2021).
\newblock Volatility analysis with realized garch-it{\^o} models.
\newblock {\em Journal of Econometrics}, 222(1):393--410.

\bibitem[Sun et~al., 2020]{sun2020adaptive}
Sun, Q., Zhou, W.-X., and Fan, J. (2020).
\newblock Adaptive huber regression.
\newblock {\em Journal of the American Statistical Association},
  115(529):254--265.

\bibitem[Tauchen et~al., 1996]{tauchen1996volume}
Tauchen, G., Zhang, H., and Liu, M. (1996).
\newblock Volume, volatility, and leverage: A dynamic analysis.
\newblock {\em Journal of Econometrics}, 74(1):177--208.

\bibitem[Xiu, 2010]{xiu2010quasi}
Xiu, D. (2010).
\newblock Quasi-maximum likelihood estimation of volatility with high frequency
  data.
\newblock {\em Journal of Econometrics}, 159(1):235--250.

\bibitem[Zhang, 2006]{zhang2006efficient}
Zhang, L. (2006).
\newblock Efficient estimation of stochastic volatility using noisy
  observations: A multi-scale approach.
\newblock {\em Bernoulli}, 12(6):1019--1043.

\bibitem[Zhang, 2011]{zhang2011estimating}
Zhang, L. (2011).
\newblock Estimating covariation: Epps effect, microstructure noise.
\newblock {\em Journal of Econometrics}, 160(1):33--47.

\bibitem[Zhang et~al., 2016]{zhang2016jump}
Zhang, X., Kim, D., and Wang, Y. (2016).
\newblock Jump variation estimation with noisy high frequency financial data
  via wavelets.
\newblock {\em Econometrics}, 4(3):34.

\end{thebibliography}

\newpage
\appendix
\section{Appendix}\label{Appendix}

\subsection{Proof of Proposition \ref{prop-integratedVol}}
  \textbf{Proof of Proposition \ref{prop-integratedVol}.} 
  We have
\begin{eqnarray*}
	V_n &=& \int_{n-1}^{n} \sigma_{t} ^{2}dt + \int_{n-1}^n J_t^2 d L_t  \cr
		&=& \int_{n-1}^{n} \sigma_{t} ^{2}dt   + E \( J_t^2 \) \lambda  +  \int_{n-1}^n \{ J_t ^2 - E \( J_t^2 \) \}  d L_t     + \int_{n-1}^n E \( J_t^2 \) ( d L_t - \lambda dt) \text{ a.s.}
\end{eqnarray*}
  We investigate $ \int_{n-1}^{n} \sigma_{t} ^{2}dt $.   
  For $k,n \in \mathbb{N}$, let
	\begin{equation*}
	R(k)\equiv  \int_{n-1}^{n}\frac{(n-t)^{k}}{k!}\sigma_{t}^{2} dt.
	\end{equation*}
By It\^o's Lemma, we have
	\begin{align*}
	R(k)= 
	& \frac{\gamma  \omega_1 -\omega_2  +  \gamma     \sigma_{n-1}^{2}    }{(k+1)!}  +\frac{ \omega_2 - 2 \gamma \omega_1 + (1-2\gamma) \sigma_{n-1}^{2}     }{(k+2) k!} \\
	&  +\frac{\gamma \omega_1 + \gamma \sigma_{n-1}^{2}    }{(k+3)k!}  - \alpha  \frac{  \lambda \omega_L + \mu  }{(k+2)!}  +\beta R(k+1) \\
	&-\alpha  \int_{n-1}^{n}  \frac{(n-t)^{k+1}}{(k+1)!} ( J_t-  \omega_L ) d  L_t - \alpha \int_{n-1}^n\frac{(n-t)^{k+1}}{(k+1)!} \omega_L (d L_t - \lambda dt)  \\
	&- \alpha  \int_{n-1}^{n}  \frac{(n-t)^{k+1}}{(k+1)!} \sigma_t dB_t +    \int_{n-1}^{n}  \frac{(n-t)^{k+1}}{(k+1)!} \varphi_t dW_t .
  \end{align*}
Then, simple algebraic manipulations show  
	\begin{align*}
	&\int_{n-1}^{n} \sigma_{t} ^{2}dt = R(0) \\
	&=   -\varrho_2\alpha (   \lambda \omega_L  +  \mu) + 2 \varrho_3 \gamma \omega_1 - \varrho_2 \omega_2 + (\varrho_1- \varrho_2 + 2\gamma \varrho_3)  \sigma_{n-1}^{2}    +D_{1,n}^J+ D_n ^c \quad \text{a.s.},
	\end{align*}
	where 
				\begin{eqnarray*}
 D_{1,n}^J = - \alpha  \beta^{-1} \left \{   \int_{n-1}^{n}  \(e ^{\beta (n-t)} -1 \) (J_t - \omega_L)  dL_t  + \omega_L \int_{n-1}^{n}  \(e ^{\beta (n-t)} -1 \) (d L_t -\lambda dt )  \right \}.
			\end{eqnarray*}
Since
\begin{equation*}
	\sigma_n ^2   =  \omega+ \gamma \sigma_{n-1}^2 +  \beta \int_{n-1}^n \sigma_s ^2 ds  - \alpha (X_n - X_{n-1}), 
\end{equation*}
we have
	\begin{align*}
	h_{n} ^c (\theta) 
	&=   -  \varrho_2  \alpha(\lambda \omega_L  + \mu) + 2 \varrho_3 \gamma \omega_1 - \varrho_2 \omega_2 + (\varrho_1- \varrho_2 + 2\gamma \varrho_3)  \sigma_{n-1}^{2}    \\
	&= -  \varrho_2  \alpha (  \lambda \omega_L  + \mu) + 2 \varrho_3 \gamma \omega_1 - \varrho_2 \omega_2 \\
	&\quad   + (\varrho_1- \varrho_2 + 2\gamma \varrho_3)   \left( \omega+ \gamma \sigma_{n-2}^2 +  \beta \int_{n-2}^{n-1} \sigma_s ^2 ds - \alpha (X_{n-1} - X_{n-2})  \right)   \\
	&=  \omega_1^g + \gamma h_{n-1}^c  (\theta)   + \beta^g \int_{n-2}^{n-1} \sigma^2_{s}   ds    - \alpha^g (X_{n-1} - X_{n-2}) ,
	\end{align*}
where  $\omega_1^g = \gamma (\rho_1 -\varrho_2 + 2 \varrho_3) \omega_1 - ( \varrho_1 -  \gamma \varrho_2 +2\gamma \varrho_3 ) \omega_2 -  (1-\gamma)      \varrho_2  \alpha (\lambda \omega_L + \mu) $.
Thus, we have
	\begin{equation*}
	\int_{n-1}^{n} \sigma_{t} ^{2}dt = h_{n} (\theta)  + D_n \text{ a.s.},
	\end{equation*}
where $D_n = D_n^c + D_n^J$. 
  \endpf

\subsection{Proof of Theorem \ref{Thm-theta}}

Define
	\begin{align*}
	\hat{L}_{n,m}   ( \theta) =&  -\frac{1}{2n}  \sum_{i=1}^n  \ell_{\tau_n}  ( \hat{V}_i - \hat{h} _ i (\theta)  )\quad \text{and} \quad \hat{\psi}_{n,m} (\theta) =\frac{\partial \hat{L}_{n,m} (\theta)}{\partial \theta}; \\
	\hat{L}_{n}  ( \theta) =&  -\frac{1}{2n}  \sum_{i=1}^n  \ell_{\tau_n}  (\hat{V}_i - h  _ i  (\theta)   )  \quad \text{and} \quad \hat{\psi}_n (\theta)=\frac{\partial \hat{L}_{n} (\theta) }{\partial \theta}; \\
	L_{n} (\theta) =&  -\frac{1}{2n}  \sum_{i=1}^n  E \[ \ell_{\tau_n}  (\hat{ V}_i - h  _ i  (\theta) ) \middle | \FF_{i-1} \] \quad \text{and} \quad \psi_n (\theta  )=\frac{\partial L_{n} (\theta)}{\partial \theta}.
	\end{align*}
To ease notations, we denote derivatives of any given function $g$ at $x_0$ by
	\begin{equation*}
	\frac{\partial g(x_0)}{\partial x} = \frac{\partial g(x)}{\partial x} \Bigg|_{x=x_0}.
	\end{equation*}
Lemma 1 in \citet{kim2016unified} shows that the dependence of $h_i(\theta)$ on the initial value decays exponentially.
Thus, we may use the true initial value $\sigma^2_0 $ during the rest of the proofs.

\begin{lemma}
	\label{Lemma-L}
	Under Assumption \ref{assumption1}(a)-(e), we have
	\begin{eqnarray}
	 && \sup_{\theta \in \Theta} \left |\hat{L}_{n,m} (\theta)-\hat{L}_n  (\theta) \right | = O_p(m^{-1/4}),  \label{Lemma-L-result1} \\
	 && \sup_{\theta \in \Theta} \tau_n ^{b-2}  \left | \hat{L}_{n} (\theta)-L_n (\theta) \right | =o_p(1), \label{Lemma-L-result2} \\ 
	&&\sup_{\theta \in \Theta}  \tau_n ^{b-2}  \left |\hat{L}_{n,m} (\theta)-L_n (\theta) \right | = O_p( \tau_n ^{b-2}  m^{-1/4}) +o_p(1).  \label{Lemma-L-result3}
	\end{eqnarray}
\end{lemma}
\textbf{Proof of Lemma \ref{Lemma-L}}. 
First, consider \eqref{Lemma-L-result1}.
By the compactness of $\Theta$, we can show
\begin{eqnarray*}
 E \[ \sup_{\theta \in \Theta} \left |   \frac{\partial \ell_{\tau_n}  (\hat{V}_i - h_i   (\theta)   ) }{\partial x}  \right | \]     \leq C   \quad \text{and} \quad  E \[ \sup_{\theta  \in \Theta} \left |  \frac{\partial ^2 \ell_{\tau_n}  (\hat{V}_i- h_i   (\theta)   ) }{(\partial x)^2}  \right | \]   \leq C.
\end{eqnarray*}
Thus, by Assumption \ref{assumption1}(d), we have
\begin{eqnarray*}
&&\sup_{\theta \in \Theta} \left | \hat{L}_{n,m} (\theta)-\hat{L}_n  (\theta) \right | \cr
 &&\leq  \frac{C}{n} \sup_{\theta \in \Theta}   \Bigg ( \left |  \sum_{i=1}^n  \frac{\partial \ell_{\tau_n}  (\hat{V}_i- h_i  (\theta)   ) }{\partial x}   \{ \hat{h} _ i   (\theta) -   h  _ i   (\theta)\}    \right |   \cr
	&& \qquad \qquad \qquad \qquad     + \left |  \sum_{i=1}^n  \frac{\partial ^2 \ell_{\tau_n}  (\hat{V}_i- h_i (\theta)   ) }{(\partial x)^2} \{\hat{h} _ i   (\theta) -   h  _ i   (\theta)\}^2  \right | \Bigg )  \cr
	&&= O_p ( m^{-{1/4}}).
\end{eqnarray*}

Consider \eqref{Lemma-L-result2}.
By the application of Theorem 2.22 in \citet{hall2014martingale} with the uniform integrability, we can show
 \begin{equation*}
	\tau_n ^{b-2}  \{ \hat{L}_n (\theta) - L_n (\theta)\} \overset { p} {\to} 0 . 
\end{equation*}
Now, we will show its uniform convergence.
Define
$$G_n(\theta) =\tau_n ^{b-2} \{ \hat{L}_n (\theta) - L_n (\theta)\}.$$
If $G_n(\theta)$ is stochastic equicontinuous, Theorem 3 in \citet{andrews1992generic} implies that it uniformly converges to zero.
Thus, it is enough to show the stochastic equicontinuity of $G_n(\theta)$.
By the Taylor expansion, we have
	\begin{eqnarray*}
	&&\left |G_n(\theta) -G_n(\theta')\right |  \cr
	&&\leq   C \frac{\tau_n^{b-2} }{2n} \sum_{i=1}^{n}  \Bigg ( \left \| \sup_{\theta^* \in \Theta} \frac{\partial \ell_{\tau_n}  ( \hat{V}_{i}- h_i(\theta^*)   ) }{\partial x}  \frac{\partial   h_i  (\theta^*)   }{\partial  \theta}  \right \|_{\max}  \cr
	&& \qquad  + \left  \|\sup_{\theta^* \in \Theta}  E \[   \frac{\partial \ell_{\tau_n}  (\hat{V}_{i}- h_i(\theta^*)   ) }{\partial x} \middle | \FF_{i-1} \]   \frac{ \partial  h_i  (\theta^*)   }{\partial  \theta}  \right \|_{\max} \Bigg )  \left \|  \theta -\theta '  \right \|_{\max}.
	\end{eqnarray*}
	For any random variables which have the finite $b$-th moment, we have
\begin{eqnarray*}
	&&E \[| \frac{\partial \ell_{\tau_n}  (x) }{\partial x}  y  |\]  \cr
	&&\leq  	2 E \[   | x y|  \1 _{\{|x| \leq \tau_n  \} }    + \tau_n |y|  \1 _{\{|x| \geq \tau_n  \} }  \]  \cr
		&&\leq 2 \tau_n E \[   |y| ^b \] ^{1/b}  E \[   \( \frac{|x|}{\tau_n}    \1 _{\{|x| \leq \tau_n  \} } \)^{b/(b-1)}  \]^{1-1/b} +   2 \tau_n E\[ |y|^b\] ^{1/b}  E\[  \1 _{\{|x| \geq \tau_n  \} }  \] ^{1-1/b} \cr
		&& \leq  C \tau_n E \[   \( \frac{|x|}{\tau_n}    \1 _{\{|x| \leq \tau_n  \} } \)^{b }  \]^{1-1/b} +   C \tau_n ^{2-b}   \cr
		&& \leq  C \tau_n ^{2-b} , 
\end{eqnarray*}
where the second and third inequalities are due to the H\"older's inequality and Markov's inequality, respectively. 
Thus, we have
\begin{eqnarray*}
	E \[ \left \| \sup_{\theta^* \in \Theta} \frac{\partial \ell_{\tau_n}  ( \hat{V}_{i}- h_i(\theta^*)  ) }{\partial x}  \frac{ \partial   h_i   (\theta^*)   }{\partial  \theta}  \right \|_{\max} \] \leq C \tau_n ^{2-b}.
\end{eqnarray*}
Similarly, using the fact that 
\begin{eqnarray*}
	 E \[   \frac{\partial \ell_{\tau_n}  (\hat{V}_{i}- h_i(\theta^*)   ) }{\partial x} \middle | \FF_{i-1} \]
\leq   E \[   \frac{\partial \ell_{\tau_n}  (\tilde{D}_i  ) }{\partial x} \middle | \FF_{i-1} \]+  \frac{\partial \ell_{\tau_n}  (h_i(\theta_0)- h_i(\theta^*)   ) }{\partial x} \text{ a.s.},
\end{eqnarray*}
we can show
\begin{eqnarray*}
	E \[ \left  \|\sup_{\theta^* \in \Theta}  E \[   \frac{\partial \ell_{\tau_n}  (\hat{V}_{i}- h_i(\theta^*)   ) }{\partial x} \middle | \FF_{i-1} \]   \frac{ \partial  h_i  (\theta^*)   }{\partial  \theta}  \right \|_{\max} \] \leq C \tau_n ^{2-b}.
\end{eqnarray*}
Thus, $G_n(\theta)$ is stochastic equicontinuous. 
\eqref{Lemma-L-result3} is the immediate result from \eqref{Lemma-L-result1} and \eqref{Lemma-L-result2}.
\endpf

\begin{lemma}
	\label{Lemma-theta}
	Under Assumption \ref{assumption1}(a)-(f), let $\theta_{\tau}$ be the maximizer of $L_{n} (\theta)$.
	Then, we have for any $u>0$,
	\begin{equation}\label{eq_Lemma-theta}
	\Pr \left\{  \left\|  \theta_{\tau} - \theta_0 \right\| _{2} \leq C\tau_n ^{1-b+u}   \right\}= 1-o(1).
	\end{equation}
\end{lemma}

\textbf{Proof of Lemma \ref{Lemma-theta}.}
It is enough to show the statement for $0<u<b-1$.
By the optimality of $\theta_{\tau}$ and the integral form of the Taylor's expansion, we have
\begin{eqnarray}\label{taylor-lemma-theta}
0 &\geq& L_{n} (\theta_{0}) - L_{n} (\theta_{\tau})\cr
&=& \langle  \psi_n (\theta_0) , \theta_{0}-\theta_{\tau} \rangle \cr
&& - \int^{1}_{0} \left(1-k\right)(\theta_{\tau}-\theta_{0})^{\top}\nabla\psi_n (\theta_{0}+k(\theta_{\tau}-\theta_{0}))(\theta_{\tau}-\theta_{0}) dk.
\end{eqnarray}
Consider $ \psi_n (\theta_0)$. Since $E \[ \tilde{ D}_i \middle | \FF_{i-1} \] =0$ a.s., we have
\begin{eqnarray*}  
  \psi_n (\theta_0)    =  \frac{1}{n} \sum_{i=1}^n   E \[  -   \tilde{D}_i    \1_{\{ | \tilde{D}_i   | \geq \tau_n \} }  +   \tau_n  \1_{\{  \tilde{D}_i   \geq \tau_n \} }   -  \tau_n  \1_{\{\tilde{D}_i   \leq  -\tau_n \} }   \middle| \FF_{i-1}\]   \frac{\partial h_i   (\theta_0) }{\partial \theta} \text{ a.s.} 
\end{eqnarray*}
Then, we have
\begin{eqnarray*}
&&E \[  \left |  \tilde{D}_i \right |  \1_{\{ |\tilde{D}_i | \geq \tau_n \}} +  \tau_n  \1_{\{  |\tilde{D}_i|   \geq \tau_n \} }   \middle | \FF_{i-1}     \]  \cr
&&\leq   E \[ |\tilde{D}_i |^b   \middle | \FF_{i-1}     \] ^{1/b} E \[ \1_{\{ |\tilde{D}_i | \geq \tau_n \} }  \middle | \FF_{i-1}      \] ^{1- 1/b} + \tau_n  E \[ \1_{\{ |\tilde{D}_i | \geq \tau_n \} }  \middle | \FF_{i-1}      \]  \cr
&&\leq  C \tau_n ^{1-b}  \text{ a.s.},
\end{eqnarray*}
where the first and second inequalities are due to H\"older's inequality and Markov's inequality, respectively. 
Thus, for any $u>0$, we have
\begin{equation} \label{eq1-lemma-theta}
\Pr \left\{  \left\|  \psi_n (\theta_0) \right\| _{2} \leq \tau_n ^{1-b+u}   \right\}= 1-o(1).
\end{equation}

Consider $-\triangledown  \psi_n (\theta)$. We have
 \begin{eqnarray} \label{eq2-lemma-theta}
 	 &&\triangledown \psi_n (\theta )  \cr
 	 &&=  \frac{1}{n} \sum_{i=1}^n   E \[   (\hat{V}_i - h  _ i (\theta))    \1_{\{ | \hat{V}_i - h  _ i (\theta)  | \leq \tau_n \} }  +   \tau_n  \1_{\{ \hat{V}_i - h  _ i (\theta) \geq \tau_n \} }   -  \tau_n  \1_{\{\hat{V}_i - h  _ i (\theta)   \leq  -\tau_n \} }   \middle| \FF_{i-1}\]    \frac{\partial ^2  h_i   (\theta) }{\partial \theta \partial \theta ^{\top} } \cr
 	 && \quad - \frac{1}{n} \sum_{i=1}^n E \[    \1_{\{ |  \hat{V}_i - h  _ i (\theta)   | \leq \tau_n \} }    \middle| \FF_{i-1}\]  \frac{\partial h_i ( \theta ) }{\partial \theta} \frac{\partial h_i ( \theta ) }{\partial \theta ^{\top} } .
 \end{eqnarray}
For the first term of \eqref{eq2-lemma-theta}, we have
\begin{eqnarray*}
    &&E \[   (\hat{V}_i - h  _ i (\theta))    \1_{\{ | \hat{V}_i - h  _ i (\theta)  | \leq \tau_n \} }     \middle| \FF_{i-1}\]    \cr
    && =     E \[   (\tilde{D}_i+ h_i(\theta_0)  - h  _ i (\theta))    \1_{\{ | \hat{V}_i - h  _ i (\theta)  | \leq \tau_n \} }     \middle| \FF_{i-1}\]  \cr
    &&=   E \[   (  h_i(\theta_0)  - h  _ i (\theta))    \1_{\{ | \hat{V}_i - h  _ i (\theta)  | \leq \tau_n \} }     \middle| \FF_{i-1}\]   - E \[   \tilde{D}_i    \1_{\{ | \hat{V}_i - h  _ i (\theta)  | > \tau_n \} }     \middle| \FF_{i-1}\]  \text{ a.s.}
\end{eqnarray*}
Thus, we have
\begin{eqnarray*}
  &&\frac{1}{n} \sum_{i=1}^n   E \[   (\hat{V}_i - h  _ i (\theta))    \1_{\{ | \hat{V}_i - h  _ i (\theta)  | \leq \tau_n \} }  +   \tau_n  \1_{\{ \hat{V}_i - h  _ i (\theta) \geq \tau_n \} }   -  \tau_n  \1_{\{\hat{V}_i - h  _ i (\theta)   \leq  -\tau_n \} }   \middle| \FF_{i-1}\]   \frac{\partial ^2  h_i   (\theta) }{\partial \theta \partial \theta ^{\top} }  \cr
  &&=\frac{1}{n} \sum_{i=1}^n   E \[    \1_{\{ | \hat{V}_i - h  _ i (\theta)  | \leq \tau_n \} }     \middle| \FF_{i-1}\]   (h_i(\theta_0) - h  _ i (\theta))    \frac{\partial ^2  h_i   (\theta) }{\partial \theta \partial \theta ^{\top} }  + O_p (\tau_n^{1-b}),
\end{eqnarray*}
where the last equality is due to the Markov's inequality.
For the second term of \eqref{eq2-lemma-theta}, we have
\begin{eqnarray*}
	 &&\frac{1}{n} \sum_{i=1}^n E \[    \1_{\{ |  \hat{V}_i - h  _ i (\theta)   | \leq \tau_n \} }    \middle| \FF_{i-1}\]  \frac{\partial h_i ( \theta ) }{\partial \theta} \frac{\partial h_i ( \theta ) }{\partial \theta ^{\top} }  \cr
	 &&=\frac{1}{n} \sum_{i=1}^n    \frac{\partial h_i ( \theta ) }{\partial \theta} \frac{\partial h_i ( \theta ) }{\partial \theta ^{\top} } -   \frac{1}{n} \sum_{i=1}^n E \[    \1_{\{ |  \hat{V}_i - h  _ i (\theta)   |  > \tau_n \} }    \middle| \FF_{i-1}\]  \frac{\partial h_i ( \theta ) }{\partial \theta} \frac{\partial h_i ( \theta ) }{\partial \theta ^{\top} }  \cr
	 &&=\frac{1}{n} \sum_{i=1}^n    \frac{\partial h_i ( \theta ) }{\partial \theta} \frac{\partial h_i ( \theta ) }{\partial \theta ^{\top} }  +  o_p (1),
\end{eqnarray*}
where the last equality is due to the Markov's inequality. 
Therefore, we have
 \begin{eqnarray*} 
 	 &&-\triangledown \psi_n ( \theta )  \cr
 	 &&=  \frac{1}{n} \sum_{i=1}^n    \frac{\partial h_i (  \theta ) }{\partial \theta} \frac{\partial h_i (  \theta ) }{\partial \theta ^{\top} }  - \frac{1}{n} \sum_{i=1}^n   E \[    \1_{\{ | \hat{V}_i - h  _ i (\theta)  | \leq \tau_n \} }     \middle| \FF_{i-1}\]   (h_i(\theta_0) - h  _ i (\theta))    \frac{\partial ^2  h_i   (\theta) }{\partial \theta \partial \theta ^{\top} } +  o_p (1).
 \end{eqnarray*}
Define 
\begin{equation*}
 \Sigma_{n}(\theta)= \frac{1}{n} \sum_{i=1}^n    \frac{\partial h_i (  \theta ) }{\partial \theta} \frac{\partial h_i (  \theta ) }{\partial \theta ^{\top} } .
\end{equation*}
Since the log stock price process is non-degenerating, there exists positive constant $0<\kappa\leq 1$ such that 
\begin{equation*}
\Pr \Big[ \inf\{ a^{\top}\Sigma_{n}(\theta_0) a: \text{ } a \in \mathbb{R}^4, \text{ } \left\| a\right\| _{2}= 1
\} \geq 2\kappa \Big] = 1-o(1).
\end{equation*}
Then, using the fact that $\tau_n ^{1-b+u}=o(1)$, we can show
\begin{equation*}
\Pr \Big[ \inf\{ -a^{\top}\triangledown \psi_n ( \theta )  a: \text{ } a \in \mathbb{R}^4, \text{ } \left\| a\right\| _{2}= 1,\text{ } \left\| \theta -\theta_{0}\right\| _{2}\leq 2\tau_n ^{1-b+u}/\kappa
\} \geq \kappa \Big] = 1-o(1).
\end{equation*}

Now, to obtain \eqref{eq_Lemma-theta}, 
it is enough to show  $\left\|  \theta_{\tau}-\theta_{0} \right\| _{2} \leq 2\tau_n ^{1-b+u}/\kappa$ under the events  
 \begin{eqnarray*} 
&&E_1= \left\{\left\|  \psi_n (\theta_0) \right\| _{2} \leq \tau_n ^{1-b+u}\right\} \text{ and } \cr
&&E_2=\left\{\inf\{ -a^{\top}\triangledown \psi_n ( \theta )  a: \text{ } a \in \mathbb{R}^4, \text{ } \left\| a\right\| _{2}= 1,\text{ } \left\| \theta -\theta_{0}\right\| _{2}\leq 2\tau_n ^{1-b+u}/\kappa\} \geq \kappa \right\}.
 \end{eqnarray*}
Suppose for the sake of contradiction that 
\begin{equation}\label{contradiction-lemma-theta}
\left\|  \theta_{\tau}-\theta_{0} \right\| _{2} > 2\tau_n ^{1-b+u}/\kappa
\text{ under } E_1 \cap E_2.
\end{equation}
Let
\begin{equation*}
z= \frac{2\tau_n ^{1-b+u}}{\kappa\left\|  \theta_{\tau}-\theta_{0} \right\| _{2}}<1.
\end{equation*}
Then for any $0\leq k \leq z$, we have
\begin{equation*}
\left\| \theta_{0} + k\left(\theta_{\tau}-\theta_{0}\right) -\theta_{0}  \right\| _{2}
\leq z\left\| \theta_{\tau}-\theta_{0} \right\| _{2} \leq 2\tau_n ^{1-b+u}/\kappa.
\end{equation*}
Thus, combining \eqref{taylor-lemma-theta} and the definitions of $E_1$ and $E_2$,  we have
 \begin{eqnarray*} 
&&\tau_n^{1-b+u}\left\| \theta_{\tau}-\theta_{0} \right\|_{2} \cr
&&\geq \int_{0}^{z}(1-k)\kappa \left\| \theta_{\tau}-\theta_{0} \right\|_{2}^{2}dk \cr
&&= 2\tau_n ^{1-b+u}\left\| \theta_{\tau}-\theta_{0} \right\|_{2} / \kappa - 2\tau_n ^{2-2b+2u}/ \kappa^{2},
\end{eqnarray*}
which contradicts to \eqref{contradiction-lemma-theta}.
\endpf


\textbf{Proof of Theorem \ref{Thm-theta}.}
By the convexity of $L_{n} (\theta)$ and Lemma \ref{Lemma-L}, we can show 
$$
\hat{\theta} \overset{p}{\to} \theta_{\tau}. 
$$ 
Furthermore, by Lemma \ref{Lemma-theta}, we have
$$
\hat{\theta} \overset{p}{\to} \theta_{0}. 
$$

By the mean value theorem and Taylor expansion, there exists $ \theta^*$ between $\theta_0 $ and $\hat{\theta} $ such that
	\begin{equation*}
	\hat{\psi} _{n,m}   (\theta _0 )- \hat{\psi} _{n,m}   (\hat{\theta} ) =  \hat{\psi} _{n,m}  (\theta _0 )=- \triangledown \hat{\psi} _{n,m}  ( \theta^*) (\hat{\theta}  -\theta _0 ).
	\end{equation*}
	We first consider $- \triangledown \hat{\psi} _{n,m}  ( \theta^*)$. 
We have
\begin{eqnarray}\label{eq00001}
-\triangledown \hat{\psi} _{n,m}  (\theta^*) &=&  \frac{1}{2n}  \sum_{i=1}^n  \ell_{\tau_n} ^{\prime \prime} ( \hat{V}_i - \hat{h} _ i (\theta^*)  )  \frac{\partial \hat{h}_i (\theta^*) } { \partial \theta}   \frac{\partial \hat{h}_i (\theta^*) } { \partial \theta ^{\top} }  \cr
&& - \frac{1}{2n}  \sum_{i=1}^n  \ell_{\tau_n} ^{\prime} ( \hat{V}_i - \hat{h} _ i (\theta^*)  )  \frac{\partial  ^2 \hat{h}_i (\theta^*) } { \partial \theta \partial \theta ^{\top} } .
\end{eqnarray}		
For the second term in \eqref{eq00001}, by Assumption \ref{assumption1}(d) and the consistency of $\hat{\theta}$, we have
\begin{equation*}
  \ell_{\tau_n} ^{\prime} ( \hat{V}_i - \hat{h} _ i (\theta^*)  )   =   \ell_{\tau_n} ^{\prime} ( \hat{V}_i -h_ i (\theta _0)  )  + o_p(1).
\end{equation*}
 Then, similar to the proofs of Lemma \ref{Lemma-theta}, we can show 
 \begin{eqnarray*}
 	 \frac{1}{2n}  \sum_{i=1}^n  \ell_{\tau_n} ^{\prime} ( \hat{V}_i - \hat{h} _ i (\theta^*)  )  \frac{\partial  ^2 \hat{h}_i (\theta^*) } { \partial \theta \partial \theta ^{\top} } &=& 	 \frac{1}{2n}  \sum_{i=1}^n  \ell_{\tau_n} ^{\prime} ( \hat{V}_i - h _ i (\theta_0)  )  \frac{\partial  ^2 \hat{h}_i (\theta^*) } { \partial \theta \partial \theta ^{\top} }  +  o_p(1) \cr
 	 &=&	    o_p(1).
 \end{eqnarray*} 
 For the first term in \eqref{eq00001}, we have
 \begin{eqnarray*}
  &&\frac{1}{2n}  \sum_{i=1}^n  \ell_{\tau_n} ^{\prime \prime} ( \hat{V}_i - \hat{h} _ i (\theta^*)  )  \frac{\partial \hat{h}_i (\theta^*) } { \partial \theta}   \frac{\partial \hat{h}_i (\theta^*) } { \partial \theta ^{\top} } \cr
  && =  \frac{1}{n}  \sum_{i=1}^n    \frac{\partial \hat{h}_i (\theta^*) } { \partial \theta}   \frac{\partial \hat{h}_i (\theta^*) } { \partial \theta ^{\top} } -  \frac{1}{n}  \sum_{i=1}^n \1 _{\{ |\hat{V}_i - \hat{h} _ i (\theta^*) | \geq \tau_n \}}   \frac{\partial \hat{h}_i (\theta^*) } { \partial \theta}   \frac{\partial \hat{h}_i (\theta^*) } { \partial \theta ^{\top} } \cr
  &&=   \frac{1}{n}  \sum_{i=1}^n    \frac{\partial \hat{h}_i (\theta^*) } { \partial \theta}   \frac{\partial \hat{h}_i (\theta^*) } { \partial \theta ^{\top} } + o_p(1),
 \end{eqnarray*}
 where the last equality is due to the Markov's inequality. By Assumption \ref{assumption1}(d) and the consistency of $\hat{\theta}$, we have
 \begin{eqnarray*}
  \frac{1}{n}  \sum_{i=1}^n    \frac{\partial \hat{h}_i (\theta^*) } { \partial \theta}   \frac{\partial \hat{h}_i (\theta^*) } { \partial \theta ^{\top} }  =  \frac{1}{n}  \sum_{i=1}^n    \frac{\partial  h _i (\theta_0) } { \partial \theta}   \frac{\partial  h_i (\theta_0) } { \partial \theta ^{\top} }  + o_p (1).
 \end{eqnarray*}
 We have  almost surely
  \begin{eqnarray*}  		 
 			 h_n   (\theta) &=& \omega^g + \gamma h_{n-1}   (\theta)   + \beta^g \int_{n-2}^{n-1} \sigma^2_{s}   ds   - \alpha^g (X_{n-1} - X_{n-2}) \cr
 				&=&  \frac{\omega^g}{ 1- \beta^g - \gamma} + \sum_{k=0}^{\infty} ( \beta^g + \gamma)^k  \left \{  \beta^g D_{n-1-k} -  \beta^g \int_{n-2-k}^{n-1-k} J_s^2 dL_s  - \alpha^g (X_{n-1-k} - X_{n-2-k})  \right \}. 
 \end{eqnarray*}
 Thus, by Assumption \ref{assumption1}(f), $h_n   (\theta_0)$ and $\int_{n-1}^{n} \sigma^2_{s}   ds$ are stationary ergodic processes. 
Therefore, by the ergodic  convergence theorem, we have
 \begin{eqnarray}\label{eq-r1}
 -\triangledown \hat{\psi} _{n,m}  (\theta^*) &=&     \frac{1}{n}  \sum_{i=1}^n    \frac{\partial  h _i (\theta_0) } { \partial \theta}   \frac{\partial  h_i (\theta_0) } { \partial \theta ^{\top} } + o_p(1) \cr
 	&\overset{p}{\to} &  V_2 ,
 \end{eqnarray}
 which is positive definite.

Consider $\hat{\psi} _{n,m}  (\theta _0) $. 
 We have 
	\begin{equation*}
	\hat{\psi} _{n,m}  (\theta _0) =   \{\hat{\psi} _{n,m}  (\theta _0) -   \hat{\psi} _{n}  (\theta _0)   \}   +\{  \hat{\psi} _{n}  (\theta _0) -   \psi  _{n}  (\theta _0) \}    + \psi  _{n}  (\theta _0) . 
	\end{equation*}
	For $\hat{\psi} _{n,m}  (\theta _0) -   \hat{\psi} _{n}  (\theta _0)$, by the Taylor's expansion, we have
	\begin{eqnarray*}
	 && \| \hat{\psi} _{n,m}  (\theta _0) -   \hat{\psi} _{n}  (\theta _0)\|_{\max}  \cr
	 && \leq  \frac{1}{n} \sum_{i=1}^n \left \|  \left \{ \ell_{\tau_n}^{\prime}  (\hat{V}_i- \hat{h}_i (\theta_0)   )  - \ell_{\tau_n}^{\prime}  (\hat{V}_i- h_i (\theta_0)   ) \right\} \frac{\partial h _{i} ( \theta_0) }{\partial \theta } \right \|_{\max} \cr
	 &&\quad  + \frac{1}{n} \sum_{i=1}^n \left \|  \ell_{\tau_n}^{\prime}  (\hat{V}_i- \hat{h}_i (\theta_0)   )   \left \{  \frac{\partial \hat{h} _{i} ( \theta_0) }{\partial \theta } - \frac{\partial h _{i} ( \theta_0) }{\partial \theta } \right \} \right \|_{\max}  \cr
	 &&\leq    \frac{1}{n}  \sum_{i=1}^n   \left \|  \ell_{\tau_n}^{\prime \prime}  (\hat{V}_i- h_i (\theta_0)   )  \{\hat{h} _ i   (\theta_0) -   h  _ i   (\theta_0)\}  \frac{\partial h _{i} ( \theta_0) }{\partial \theta }  \right \|_{\max}  \cr
	 	 &&\quad  + \frac{1}{n} \sum_{i=1}^n \left \|  \ell_{\tau_n}^{\prime}  (\hat{V}_i- \hat{h}_i (\theta_0)   )   \left \{  \frac{\partial \hat{h} _{i} ( \theta_0) }{\partial \theta } - \frac{\partial h _{i} ( \theta_0) }{\partial \theta } \right \} \right \|_{\max}  \cr
	 	 &&= O_p(m^{-1/4}),
	\end{eqnarray*}
	where the last equality is due to Assumption \ref{assumption1}(d).
	For $\psi  _{n}  (\theta _0)$, by the Ergodic convergence theorem, we have 
	\begin{eqnarray*}  
	&&  n^{(b-1)/b}  \psi_n (\theta_0) \cr
		&&  =  n^{(b-1)/b} \frac{1}{n} \sum_{i=1}^n   E \[  -   \tilde{D}_i    \1_{\{ | \tilde{D}_i   | \geq \tau_n \} }  +   \tau_n  \1_{\{  \tilde{D}_i   \geq \tau_n \} }   -  \tau_n  \1_{\{\tilde{D}_i   \leq  -\tau_n \} }   \middle| \FF_{i-1}\]   \frac{\partial h_i   (\theta_0) }{\partial \theta} \cr
		&& \overset{p}{\to}   S.
\end{eqnarray*}

Consider  $ \hat{\psi} _{n}  (\theta _0) -   \psi  _{n}  (\theta _0)$.
For any given $a \in \mathbb{R}^4$,  let 
	\begin{eqnarray*}
		A_i &=&   n^{(b-2)/2b}    \Big \{  \tilde{D}_i    \1_{\{ | \tilde{D}_i   | <  \tau_n \} }  +   \tau_n  \1_{\{  \tilde{D}_i   \geq \tau_n \} }   -  \tau_n  \1_{\{\tilde{D}_i   \leq  -\tau_n \} } \cr
		&&  \qquad \qquad \qquad  - E \[     \tilde{D}_i    \1_{\{ | \tilde{D}_i   | < \tau_n \} }  +   \tau_n  \1_{\{  \tilde{D}_i   \geq \tau_n \} }   -  \tau_n  \1_{\{\tilde{D}_i   \leq  -\tau_n \} }   \middle| \FF_{i-1}\]  \Big \} a^{\top} \frac{\partial h_i   (\theta_0) }{\partial \theta}.
	\end{eqnarray*}
Then, $A_i$ is a martingale difference. 
We have
\begin{equation*}
	E \[ A_i^2 \middle | \FF_{i-1} \] =  n^{(b-2)/b}   E \[     \tilde{D}_i ^2     \1_{\{ | \tilde{D}_i |  <  \tau_n \} }     +    \tau_n   ^2   \1_{\{| \tilde{D}_i |  \geq  \tau_n \}}       \middle | \FF_{i-1} \]  \(a^{\top} \frac{\partial h_i   (\theta_0) }{\partial \theta} \) ^2 \text{ a.s.}
\end{equation*}
	By the Ergodic convergence theorem, we have
	\begin{equation*}
	\frac{1}{n}  \sum_{i=1}^n E \[ A_i^2 \middle | \FF_{i-1} \]  \overset{p} {\to}  V_1 a^{\top} V_2 a.
	\end{equation*}
Thus, with \eqref{eq-r1},  the statement \eqref{Thm-theta-r1} is showed. 
	To investigate its asymptotic normality, we check the Lindeberg condition. 	
For any given $\delta >0$, we have
	\begin{eqnarray*}
		\frac{1}{n} \sum_{i=1}^n E \[ A_i ^2 \1 _{ \{ |A_i | \leq \delta n \} } \]  &\leq& \frac{1}{n} \sum_{i=1}^n E \[ A_i ^{2b} \] ^{1/b} E \[  \1 _{ \{ |A_i | \leq \delta n \} } \]  ^{1-1/b} \cr
		&\leq& \frac{1}{n} \sum_{i=1}^n E \[ A_i ^{2b}\] ^{1/b}  \(\frac{E \[ |A_i |^2 \] } {  \delta^2 n^2 } \)   ^{1-1/b} \cr
		&\leq& C \tau_n   n^{-2 + 2/b} \cr
		& =& C n^{3/b-2} \to 0 \text{ as } n \to \infty ,
	\end{eqnarray*}
	where the first inequality is due to the H\"older's inequality and the last inequality is due to \eqref{eq00} below. 	
	We have
	\begin{eqnarray}\label{eq00}
E \[ A_i ^{2b}\]  &\leq& C E \[ E \[ | \tilde{D}_i | ^{2b}  1 _{\{ |\tilde{D}_i | \leq \tau_n \}}  + \tau_n ^{2b}  1 _{\{ |\tilde{D}_i | \geq \tau_n \}} \middle| \FF_{i-1}  \]   \(a^{\top} \frac{\partial h_i   (\theta_0) }{\partial \theta} \) ^{2b} \] \cr
&\leq&C E \[  \( \tau_n ^{2b} E \[ \left |\frac{ \tilde{D}_i}{\tau_n}  \right | ^{2b} 1 _{\{ |\tilde{D}_i | \leq \tau_n \}} \middle | \FF_{i-1} \]   + \tau_n ^{b}  \)   \(a^{\top} \frac{\partial h_i   (\theta_0) }{\partial \theta} \) ^{2b} \]  \cr
&\leq&CE \[  \( \tau_n ^{2b} E \[ \left |\frac{ \tilde{D}_i}{\tau_n}  \right | ^{b} 1 _{\{ |\tilde{D}_i | \leq \tau_n \}} \middle | \FF_{i-1} \]   + \tau_n ^{b}  \)   \(a^{\top} \frac{\partial h_i   (\theta_0) }{\partial \theta} \) ^{2b} \]  \cr
&\leq&C \tau_n ^{b} E \[    \(a^{\top} \frac{\partial h_i   (\theta_0) }{\partial \theta} \) ^{2b} \],
	\end{eqnarray}
	where the second inequality is due to the Markov's inequality. 
	Thus, the Lindeberg condition is satisfied. 
By Theorem 1 \citep{brown1971martingale}, we have
\begin{eqnarray*}
n^{(b-2)/2b}   \sqrt{n}    \( \hat{\psi} _{n}  (\theta _0) -   \psi  _{n}  (\theta _0) \) \overset{d} {\to} N(0,  V_1 V_2 ).
\end{eqnarray*}
 Therefore, by the Slutsky's theorem, we have
 \begin{equation*}
 n^{(b-2)/2b}   \sqrt{n}   ( \hat{\theta} - \theta_0) \overset{d}{\to} N( V_2 ^{-1} S,  V_1 V_2 ^{-1} ).
 \end{equation*}
 \endpf

\subsection{Proof of Theorem \ref{Thm-adj}}

\textbf{Proof of Theorem \ref{Thm-adj}.}
We have
\begin{eqnarray*}
\hat{\theta}_ {adj} -\theta_0 &=& \hat{\theta}-\theta_0  +   \[     \sum_{i=1}^n   \frac{\partial \hat{h}_i ( \hat{ \theta} ) }{\partial \theta} \frac{\partial \hat{h}_i ( \hat{\theta} ) }{\partial \theta ^{\top} } \] ^{-1}   \sum_{i=1}^n   \frac{\partial \hat{h}_i   (\hat{\theta}) }{\partial \theta} (\tilde{D}_i +h_i (\theta_0) - \hat{h}_i (\hat{\theta}) )  \cr
&=& \frac{ V_2 ^{-1} }{n} \sum_{i=1}^n   \frac{\partial  h_i   (\theta_0) }{\partial \theta}  \tilde{D}_i  + o_p( \hat{\theta} -\theta_0 ) +  O_p(m^{-1/4}),
\end{eqnarray*}
where the last equality is due to \eqref{eq1-Thm-adj}--\eqref{eq2-Thm-adj} below and the consistency of $\frac{1}{n}  \sum_{i=1}^n   \frac{\partial \hat{h}_i ( \hat{ \theta} ) }{\partial \theta} \frac{\partial \hat{h}_i ( \hat{\theta} ) }{\partial \theta ^{\top} }$. 
By the Taylor's expansion and mean value theorem, there exists $\theta^*$ between $\theta_0$ and $\hat{\theta}$ such that
\begin{eqnarray}\label{eq1-Thm-adj}
\frac{1}{n} \sum_{i=1}^n   \frac{\partial \hat{h}_i   (\hat{\theta}) }{\partial \theta}  \tilde{D}_i &=& \frac{1}{n} \sum_{i=1}^n    \frac{\partial h_i   (\hat{\theta}) }{\partial \theta}  \tilde{D}_i  +  O_p(m^{-1/4})  \cr
	&=& \frac{1}{n} \sum_{i=1}^n    \frac{\partial h_i   (\theta_0) }{\partial \theta}  \tilde{D}_i  +  \frac{1}{n} \sum_{i=1}^n    \frac{\partial^2 h_i   (\theta^*) }{\partial \theta^2 }  \tilde{D}_i (\hat{\theta} - \theta_0)  +  O_p(m^{-1/4})  \cr
	&=& \frac{1}{n} \sum_{i=1}^n    \frac{\partial h_i   (\theta_0) }{\partial \theta}  \tilde{D}_i  + o_p (\hat{\theta} - \theta_0)  +  O_p(m^{-1/4}) ,
\end{eqnarray}
where the first equality is due to Assumption \ref{assumption1}(d) and the last equality is by the fact that by the Ergodic convergence theorem,  
$ \frac{1}{n} \sum_{i=1}^n    \frac{\partial^2 h_i   (\theta^*) }{\partial \theta^2 }  \tilde{D}_i \overset{p}{\to}0$. 
Similarly, we can show
\begin{eqnarray}\label{eq2-Thm-adj}
\frac{1}{n} \sum_{i=1}^n   \frac{\partial \hat{h}_i   (\hat{\theta}) }{\partial \theta} (h_i (\theta_0) - \hat{h}_i (\hat{\theta}) )  &=& \frac{1}{n} \sum_{i=1}^n   \frac{\partial \hat{h}_i   (\hat{\theta}) }{\partial \theta} \frac{\partial \hat{h}_i   (\theta^*) }{\partial \theta^{\top}}  (\theta_0 - \hat{\theta} )   \cr
	&=& \frac{1}{n} \sum_{i=1}^n   \frac{\partial \hat{h}_i   (\hat{\theta}) }{\partial \theta} \frac{\partial \hat{h}_i   (\hat{\theta}) }{\partial \theta^{\top}}  (\theta_0 - \hat{\theta} )  + o_p (\hat{\theta} - \theta_0) .
\end{eqnarray}
\endpf


\end{spacing}
\end{document}